\documentclass[review,authoryear]{elsarticle}

\usepackage{amssymb}
\usepackage{graphicx}
\usepackage{sistyle}
\usepackage{longtable}
\usepackage{wasysym}
\usepackage{color}
\newcommand{\deqS}{\ensuremath{D_{S}}}
\newcommand{\deqV}{\ensuremath{D_{V}}}

\newcommand{\rev}[1]{#1}

\bibpunct{(}{)}{;}{a}{}{,}

\usepackage{hyperref}
\hypersetup{
    bookmarks=true,         
    unicode=false,          
    pdftoolbar=true,        
    pdfmenubar=true,        
    pdffitwindow=false,     
    pdftitle={Physical and dynamical properties of the main belt triple asteroid (87) Sylvia},
    pdfauthor={J\'er\^ome Berthier}, 
    pdfsubject={Planetary Science}, 
    pdfkeywords={},         
    pdfnewwindow=true,      
    colorlinks=true,        
    linkcolor=gray,         
    citecolor=blue,         
    filecolor=gray,         
    urlcolor=gray           
}

\biboptions{authoryear}

\begin{document}

\begin{frontmatter}



\title{Physical and dynamical properties of the main belt triple asteroid (87) Sylvia\tnoteref{thx}}

\tnotetext[thx]{Based on observations collected at the European Southern Observatory, Paranal, Chile (089.C-0944, 087.C-0014, 385.C-0089, 085.C-0480, 077.C-0422, 074.C-0052), at the Gemini North Observatory in Hawaii, and at the W. M. Keck Observatory. The Keck observatory was made possible  by the generous financial support of the W. M. Keck Foundation.}  


\author[imcce]{J. Berthier\corref{cor}}
\ead{berthier@imcce.fr}
\author[imcce]{F. Vachier}
\author[seti,imcce]{F. Marchis}
\author[astroinst]{J. {\v D}urech}
\author[imcce]{B. Carry}

\cortext[cor]{Corresponding author}

\address[imcce]{Observatoire de Paris, CNRS UMR8028, Sorbonne Universit\'es, UPMC Univ Paris 06, IMCCE, 77 avenue Denfert Rochereau, 75014 Paris, France}
\address[seti]{Carl Sagan Center, SETI Institute, 189 Bernardo Avenue, Mountain View CA 94043, USA}
\address[astroinst]{Astronomical Institute, Faculty of Mathematics and Physics, Charles University in Prague, V Hole\v{s}ovi\v{c}k\'ach 2, 18000 Prague, Czech Republic}

\begin{abstract}
We present the analysis of high angular resolution observations of the triple asteroid (87) Sylvia collected with three 8--10 m class telescopes (Keck, VLT, Gemini North) and the Hubble Space Telescope. The moons' mutual orbits were derived individually using a purely Keplerian model. We computed the position of Romulus, the outer moon of the system, at the epoch of a recent stellar occultation which was successfully observed at less than 15 km from our predicted position, within the uncertainty of our model. The occultation data revealed that the moon, with a surface-area equivalent diameter $\deqS = 23.1 \pm 0.7$ km, is strongly elongated (axes ratio of $2.7 \pm 0.3$), significantly more than single asteroids of similar size in the main-belt. We concluded that its shape is probably affected by the tides from the primary. A new shape model of the primary was calculated combining adaptive-optics observations with this occultation and 40 archived light-curves recorded since 1978. The difference between the $J_2 = 0.024\,^{+0.016}_{-0.009}$ derived from the 3-D shape model assuming an homogeneous distribution of mass for the volume equivalent diameter \deqV = $273 \pm 10$ km primary and the null $J_2$ implied by the keplerian orbits suggests a non-homogeneous mass distribution in the asteroid's interior.
\end{abstract}


\end{frontmatter}


\section{Introduction}

The minor planet (87) Sylvia is a main belt asteroid discovered in 1866 by \cite{Pogson1866}. In the 1990s, frequency analysis of photometric observations hinted that this asteroid could be binary \citep{Prokofeva1992, Prokofeva1994, Prokofeva1995}. Its first satellite (S/2001 (87) 1, known as {\it Romulus}) was discovered in February 2001 by \cite{Brown2001} using the Keck II telescope atop Hawaii's Mauna Kea. Three years later, \cite{Marchis2005} announced the discovery of a second companion (S/2004 (87) 1, known as {\it Remus}), using the European Southern Observatory's Very Large Telescope (VLT). Sylvia became the first known triple asteroidal system. Since then, eight others have been discovered and studied \citep{Brown2005b, Bouchez2006, Ragozzine2009a, Brozovic2011a, Descamps2011, Fang2011, Marchis2010b, Marchis2013}.

Asteroid (87) Sylvia is the largest member of a collisional family born, at least, several hundreds of million years ago, more probably between 1 and 3.8 Gyr. The age of this family, for which more than 80 members have been identified among current census of asteroids, is commensurable with the evolutionary timescales of Sylvia's satellite system \citep{Vokrouhlicky2010}. Various authors estimate that the system is  dynamically very stable over a large timescale (at least one million years, see \cite{Winter2009, Vokrouhlicky2010, Fang2012}), the satellites being in a deeply stable zone, surrounded by both fast and secular chaotic regions due to mean-motion and evection resonances \citep{Frouard2012}.

Observations led to the determination of the dynamical and physical properties of the system. Asteroid (87) Sylvia is classified as a X-type asteroid \citep{Bus2002a} located in the outer main belt ($a \simeq 3.49$ AU, $e \simeq 0.09$, $i \simeq 11\arcdeg$), within the large Cybele-zone, with a volume-equivalent diameter estimated to $278 \pm 11$ km, a relatively low density of $1.31 \pm 0.15$, and a large macro-porosity estimated to $52 \pm 11$\% (\citealt{Carry2012c}, and references therein). The two moons, Remus and Romulus, with a diameter respectively estimated to $\sim$7 km and $\sim$18 km \citep[from photometry measurements,][]{Marchis2005a} or 9--12 km and 5--16 km \citep[derived by][as a free parameter of their dynamical model]{Fang2012}, orbit at a distance of $\sim$700 km and $\sim$1350 km from the primary. Finally, \cite{Fang2012} estimates dynamically that Sylvia is oblate with a $J_{2}$ value in the range 0.0985--0.1.

We report here on new results on the dynamical and physical properties of Sylvia's system based on the analysis of adaptive-optics imaging, light-curves and stellar occultation data (Section \ref{data}). We improve Sylvia's 3-D shape model and estimate its overall size (Section \ref{sylvia}). We estimate the shape and size of the outer satellite Romulus from the analysis of the latest observed stellar occultation (Section \ref{romulus}). We improve the determination of orbital parameters for the two satellites, we estimate the mass and density of Sylvia, and we examine its quadrupole term $J_{2}$ (Section \ref{orbits}). Finally, we discuss the surprising elongated shape of Romulus revealed by this stellar occultation.

\section{Observations and data}\label{data}

 \subsection{Adaptive-optics observations}

We gathered in the VOBAD database \citep{Marchis2006a} all observations, acquired by our group or already published, from February 2001 to December 2012 recorded in the near-infrared with adaptive-optics (AO) systems available on large ground-based telescopes. We use the ESO Very Large Telescope NACO imaging camera \citep{Lenzen2003, Rousset2003d} and SINFONI spectro-imaging camera \citep{Eisenhauer2003}, the Gemini North ALTAIR AO system \citep{Herriot2000} with its camera NIRI \citep{Hodapp2003}, and the NIRC2 camera on the Keck II telescope \citep{Wizinowich2000b, vanDam2004e}.
  
The AO frames were recorded in broad band filters (J, H, or K, from 1 to 2.5 $\mu$m) and were all processed in a similar manner. The basic data processing (sky subtraction, bad-pixel removal, and flat-field correction) applied on all these raw data was performed using the recommended eclipse data reduction package \citep{devillard1997}. Successive frames taken over a time span of less than 6 min, were combined into one single average image after applying an accurate shift-and-add process through the Jitter pipeline offered in the same package. Data processing with this software on such high signal-to-noise ratio data ($>$1000) is relatively straightforward. Since these data respect the Shannon's theorem, it is possible to retrieve completely the continuous signal from the knowledge of enough samples. After re-sampling each image to 1/10th of the pixel level, the centroid position on each frame can be accurately measured by a Gaussian fit. The final image is obtained by stacking the set of frames with individual shifts determined from cross-correlation. Once processed, individual images reveal the resolved shape of the primary (angular size $\sim\!0.2 \arcsec$), and, sometimes, the unresolved image of the satellites appears. We used a dedicated algorithm \citep{Hanuvs2013c} to extract the contour of the primary and to determine its photocenter, from which we measured the astrometric positions of the satellites by fitting a Moffat-Gauss source profile.

\begin{figure}
  \includegraphics[width=\textwidth]{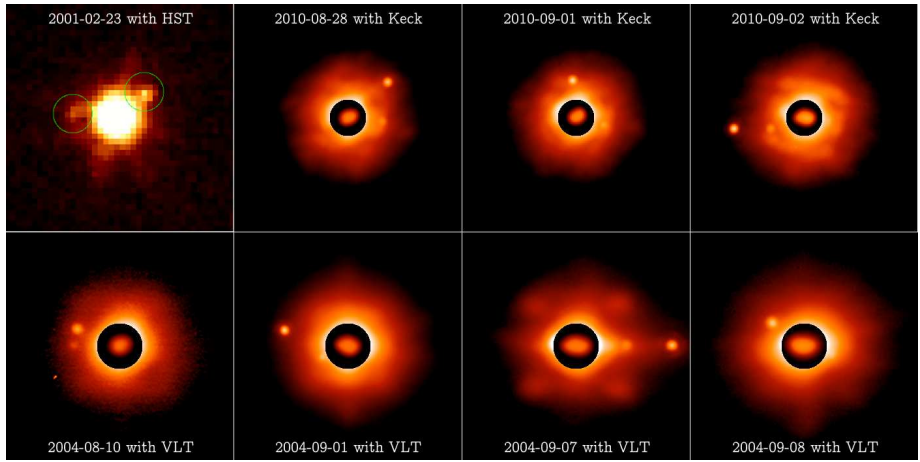}
  \caption{Sample of observations of the triple asteroid (87) Sylvia taken with the HST/WFPC2, W.M. Keck AO and VLT/NACO revealing the two satellites and the well resolved primary (angular size $\sim\!0.2\arcsec$). The high levels of intensity showing the irregular shape of the primary are shown in the central circle. North is up, and East is left. The plate scale is 40.2 mas for the HST/WFPC2, 9.9 mas for the Keck/NIRC2, and 13.3 mas for the VLT/NACO observations. \label{fig-ao}}
\end{figure}

Figure \ref{fig-ao} displays Sylvia's system as seen by VLT/NACO and Keck/NIRC2 instruments after the processing has been applied. Table \ref{tbl-observingConditions} provides the observing condition at each epoch of image acquisition. Tables \ref{tbl-obsRemus} and \ref{tbl-obsRomulus} summarize all the astrometric measurements used to fit the orbits of the two satellites of Sylvia. The accuracy on the observing time is the result of the computed mean time of the jittered images, typically 0.2s to 1s depending on the observatory where the data were recorded.


 \subsection{Hubble space telescope data}

\cite{Storrs2001} reported the confirmation of the presence of Romulus on Hubble Space Telescope (HST) images collected on February 23, 2001 with the WFPC2 instrument through various filters. We retrieved from the HST archive the three unsaturated observations taken through the F439 filter with an individual exposure time of 3s. These observations were re-processed  using our own pipeline reproducing the HST/WFPC2 cookbook method. The resulting image, shown in Fig. \ref{fig-ao}, confirmed the detection of Romulus at a position very close to the one reported by \cite{Storrs2001}. The second satellite, Remus, is also visible and detected closer to the primary at a distance of 0.34$\arcsec$. The positions of the satellite were derived using the same Moffat-Gauss profile fit than for our AO observations. Interestingly, even though \cite{Marchis2005a} reported the triple nature of (87) Sylvia from observations taken in August 2004 and onward, this February 2001 HST observation was in fact the first detection of the third component of the system. We included these astrometric positions for Romulus and Remus in our analysis, which is particularly useful for Remus since it increases the observational time span by 1264 days ($\sim$931 revolutions).

 \subsection{Light-curve data}

We used 40 light-curves observed from 1978 to 1989 published by \cite{Harris1980}, \cite{Schober1979}, \cite{Weidenschilling1987c}, \cite{Weidenschilling1990c}, \cite{Blanco1989c}, and \cite{Prokofeva1992}. The data were compiled by \cite{Lagerkvist1987} in the Uppsala Asteroid Photometric Catalog, now available through Internet (\href{http://asteroid.astro.helsinki.fi/apc}{APC}\footnote{http://asteroid.astro.helsinki.fi/apc} Web site). To this set of dense light-curves we added sparse photometry from US Naval Observatory (IAU code 689), Roque de los Muchachos Observatory, La Palma (950), and Catalina Sky Survey Observatory (703). See the works by \citet{durech2005} and \citet{hanus2013} for details on sparse photometry.

 \subsection{Stellar occultation data}\label{occdata}

The observation of a stellar occultation consists in recording the time of immersion and emersion of the star in front of which the asteroid passes, as seen by geographically distributed observers along the occultation path. Each observed occultation point is then projected onto a common plane that passes through the center of the Earth, and lies perpendicular to the direction of the star as seen from the occulting body. Assuming that the relative velocity of the body with respect to the observer is well estimated by the ephemeris, which is a soft assumption, especially for numbered asteroids, one can transform the timings into lengths, and then evaluate the size of the occulting body. We get several segments - the chords - which are directly proportional to the size of the cross-section of the body as seen by observers. With a sufficient number of chords, the silhouette of the body is drawn, and can yield a strong constraint on its 2-D profile in the occultation plane.

The observation of a stellar occultation by an asteroid is one of the few methods which can yield the size and the overall shape of the asteroid without hypothesis on its physical nature. If several independent events of the same occulting body are collected then a 3-D model of the asteroid can even be determined \citep{Drummond1989}. However, due to the low number of well-covered events for a given asteroid, full 3-D reconstruction based on stellar occultations only will always concern a small sample of asteroids. Stellar occultations have however proved to be useful to scale the convex shape models of asteroids derived by light-curve inversion \citep{vDurech2011}.

Four stellar occultations by Sylvia have been reported in the past 30 years, but only a total of four chords (three for an event, one for another, none for the two others) have been collected \citep{Dunham2012}. These occultation data are therefore useless to scale Sylvia's shape model \citep{Kaasalainen2002}. The stellar occultation by Sylvia successfully observed in early 2013 is thus the first opportunity to do so.

On January 6, 2013, about 50 European observers were mobilized to observe the occultation of the TYCHO-2 1856-00745-1 star by (87) Sylvia \citep{Berthier2013}. Among them, 19 observers have recorded a negative event (i.e., no disappearance of the star), and 13 observers have reported a successful observation of the event, providing 16 chords including 4 of the occultation by Romulus. The bad weather forecast on western Europe this night prevented other observers to record the event. Figure \ref{fig-occhords} shows the result of the observation of this stellar occultation. Table \ref{tbl-timings} presents the timings of the event recorded by observers (published on \href{http://euraster.net/results/2013/index.html#0106-87}{Euraster}\footnote{http://euraster.net/results/2013/} Web site), and table \ref{tbl-observers} lists the observers and the geodetic coordinates of the observing sites.


Stellar occultations by asteroids are usually observed by a group of observers who use different acquisition and timing devices. As a consequence, the accuracy on the measurements differs from one observer to another, and sometimes measurements can disagree owing to systematics in the calibration of the absolute timing reference. A typical example is a chord which is clearly shifted with respect to other chords nearby. The latter can then be used to estimate the offset to apply on the chord to restore its timings. However, it can be tricky to shift the timings of chords, mainly because no evident rule can be found. In such cases it is better to decrease the weight of uncertain chords with respect to accurate ones. 

For the January 6, 2013 stellar occultation by Sylvia, we are confident in the absolute timings of the chords as most timing references were insured by GPS or atomic clock time servers (see Tab. \ref{tbl-observers}). We considered only two chords as discrepant (see column ``Offset'' of Tab. \ref{tbl-timings}). The first shows a clear lateness of a few seconds with respect to the three chords immediately next to it (one north, two south). The offset we apply (-3.8s) sets the chord back in alignment with the three others. The second discrepant chord shows a slight lead on a nearby chord for which the observer can assert the time of the disappearance. The offset we apply (+0.5s) shift the chord forward in alignment with the latter. As we empirically changed the timings of these two chords, we \rev{down-weighted their timings to one thousandth of the weight affected to other chords}.

At the mean time of the occultation, January 6, 2013 01:37:06.46 UTC, the computed geocentric orbital velocity of Sylvia is 17.1 km/s. With timing uncertainties lying in the range 0.03--1\,s (with a mean value of $0.25 \pm 0.3$\,s), we can expect to measure lengths with a mean accuracy of 4 km ($\sim$1.5\% of Sylvia's size), which corresponds to $\sim$2 mas at the geocentric distance of Sylvia.

\section{Shape and size of the system components}

 \subsection{Shape and size of Sylvia}\label{sylvia}

A convex shape model of Sylvia was previously derived by \cite{Kaasalainen2002}. They used 32 light-curves observed in 8 apparitions spanned over 1978 - 1989, which allowed them to derive a unique model with ECJ2000 pole direction (71\arcdeg, +66\arcdeg). As was later shown by \cite{Marchis2006e}, this model agreed well with AO observations. It also agrees with the occultation data presented here.  

To further improve the fit and to include all data types into the modeling, we used the KOALA algorithm \citep{Carry2010a, carry2012b}. It can simultaneously fit optical light-curves, occultation data and image profiles to give a best-fit 3-D shape model represented by a non-convex polyhedron \citep{Kaasalainen2011}. Contrary to standard light-curve inversion, where light-curves are fitted by a convex shape model, the best-fit solution is harder to define in case of multiple data sources. Different types of data have to be weighted to compose the total $\chi^2$ as a measure of the goodness of fit \citep{Kaasalainen2011}. Moreover, given the low number of 2-D profiles available, the non-convex shape solution had to be regularized to suppress unrealistic small-scale fluctuations of the shape. The shape was approximated by a polyhedron described by a spherical harmonics series of the order and degree five or six. This resolution was sufficient to model details of the occultation profile, and still low enough to suppress artificial details that often appeared with higher resolution.

\begin{figure}
  \includegraphics[width=\textwidth]{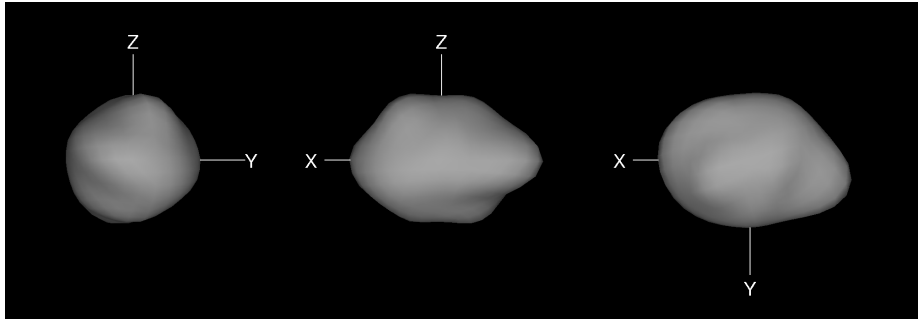}
  \caption{One of the best-fit models of Sylvia's primary (equatorial, equatorial, and polar view) derived from the combination of light-curve inversion, AO, and occultation data.\label{fig-onebestmodel}}
\end{figure}

\begin{figure}
  \includegraphics[width=\textwidth]{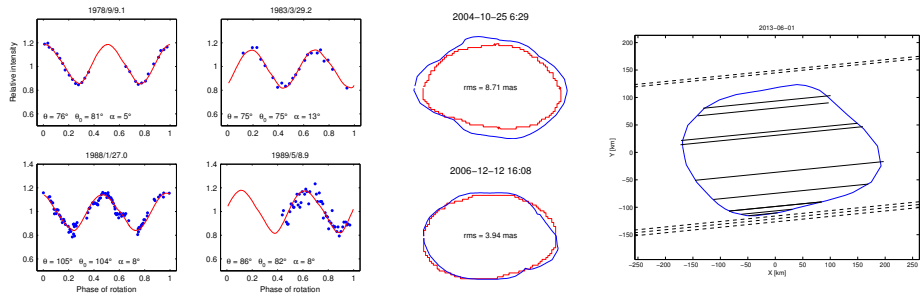}
  \caption{Result of the fit of Sylvia's 3-D shape compared to a sample of light-curves, image profiles, and occultation chords. The measured light-curves (points) are compared with the model (solid curves). The viewing and illumination geometry is described by the latitudes $\theta$ and $\theta_0$ of the sub-Earth and sub-solar point, respectively, and by the solar phase angle $\alpha$. The observed contours of AO images and the corresponding projections of the model (smooth blue curves) are plotted in the second panel. In the third panel, occultation chords and the model outline are shown. The dashed lines are negative observations.\label{fig-colfits}}
\end{figure}

\rev{The KOALA algorithm minimizes the total $\chi^2 = \chi^2_\mathrm{LC} + w_\mathrm{AO} \chi^2_\mathrm{AO} + w_\mathrm{Occ} \chi^2_\mathrm{Occ}$ that composes the individual contributions from light-curves (LC), adaptive optics profiles (AO), and occultation chords (Occ). Adaptive optics and occultation data are weighted with respect to the light-curves with parameters $w_\mathrm{AO}$ and $w_\mathrm{Occ}$, respectively. The optimum values of these weights (so called maximum compatibility weights) can be obtained following the approach of \cite{Kaasalainen2011}. In our case, the shape of the $\cal S$ surface was `shallow', thus the optimum weights were not well defined -- an order of magnitude range of $w_\mathrm{AO}$ and $w_\mathrm{Occ}$ was possible.} We varied weights of AO and occultation data with respect to light-curves to see how the models evolved, then we selected only solutions that gave acceptable fits to the data. Thus, instead of one best solution, we obtained a set of slightly different models that all fit the data well\rev{, each with a slightly different spin, shape, and size}. One of these models is shown in Fig. \ref{fig-onebestmodel}, and the fits to the light-curves, image and occultation profiles are shown in Fig. \ref{fig-colfits}. The discrepancy between one AO image contour and the model silhouette comes from an over-estimation of the AO contour due to poor seeing conditions at the time of this observations. The root mean square (RMS) residual of the occultation fit is 2.6 km, which is below the accuracy provided by the timings (see \ref{occdata}).


To \rev{further} estimate the uncertainty of our model, we created many clones of the original data set and processed them the same way as the original data. For light-curves, we re-sampled the set with random sampling and replacement -- so each new clone consisted of the same light-curves, but some were missing and some were included more times. This approach is similar to the bootstrap method \citep{Press2007}, the difference is that we re-sampled the whole light-curves, not individual measurements, because points within one light-curve cannot be treated as fully independent measurements. By this approach, we can roughly sample the range of possible solutions. For occultation data, we randomly changed the timings using a normal distribution law, with reported uncertainties as standard deviations. Because of the low weight of the AO data in the process, we left them unchanged. The distribution of shapes and spin axis directions then represented the variance of the model given the data and served for a realistic estimation of the model uncertainties. \rev{The width of $3\sigma$ error intervals reported in Table~ \ref{tbl-physParams} corresponds to the maximum spread of the relevant parameters across our sample of bootstrap models obtained with different $w_\mathrm{AO}$ and $w_\mathrm{Occ}$ weights.}

\begin{figure} 
  \includegraphics[width=\textwidth]{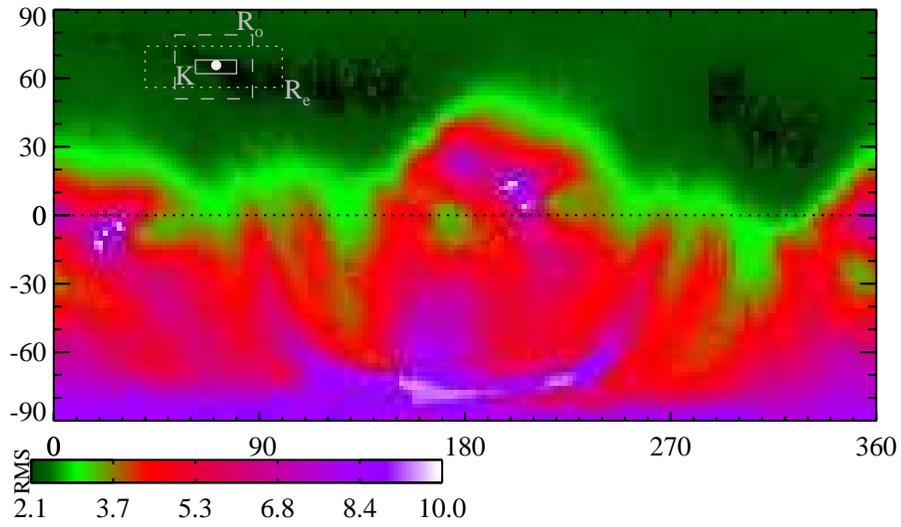}
  \caption{
    RMS residuals between the synthetic and observed lightcurves for spin-vector coordinates covering the entire celestial sphere with grid interval of 2\arcdeg. The white points shows the best-fit spin solution (70\arcdeg, +69\arcdeg), and the three rectangles correspond to the 3--$\sigma$ confidence interval provided by KOALA, the orbital pole of Romulus, and Remus, respectively, outside which spin-vectors coordinates are unlikely.\label{fig-rmspole}} 
\end{figure}

Figure \ref{fig-rmspole} displays the RMS residuals between synthetic and observed light-curves as function of the spin-vector coordinates (grid interval of 2\arcdeg on the ECJ2000 celestial sphere). For one of the formally best solutions (based on the total $\chi^2$), the direction of the spin axis (70\arcdeg, +69\arcdeg) is almost the same as for the convex model of \cite{Kaasalainen2002}, and the 3-D shapes are similar overall. The standard deviation of the pole direction is about 3\arcdeg\ in ecliptic longitude and 1\arcdeg\ in ecliptic  latitude. The volume-equivalent diameter for the best model is \deqV = 273\,km, and $J_{2}$ = 0.024. The basic physical parameters are listed in Table~\ref{tbl-physParams}. We also used different resolutions of the model (the degree and order of spherical harmonic expansion used for the shape parametrization) and different weightings between the data to assert the stability of the solution. The plausible shape models cover a range of equivalent diameters from 266 to 280\,km and $J_{2}$ spans the range 0.015\,--\,0.040.


\rev{Using our 3-D shape model, we estimated for each image the offset between the center of mass and the photocenter, assuming an uniform density distribution. They span the intervals [-2.3, +1.9] mas in $x$, and [-0.4, +1.8] mas in $y$, and are, at least, five time smaller than the uncertainties of the astrometric measurements (see Tables \ref{tbl-obsRemus} and \ref{tbl-obsRomulus}). We have not taken into account these offsets in the astrometric measurements used to fit orbits.}


 \subsection{Shape and size of Romulus}\label{romulus}

The analysis of the chords of Romulus give us the opportunity to determine its size and, for the first time, to estimate its shape. The distribution of the 4  observed chords is not sufficient to model without ambiguity the profile of the satellite, but they are sufficiently well distributed to estimate its size and biaxial dimensions. The parameters of the ellipsoidal profile (coordinates of the center $x_{0}, y_{0}$, axis radii $a, b$, and position angle from North of the major axis $\psi$) are calculated from the set of conics which best fit the $n$ points of coordinates $(x_{i}, y_{i})_{n}$, which correspond to the coordinates of the extremity of the chords as measured by the observers. We solve the system of linear equations by the linear least-squares method of the Singular Value Decomposition method \citep{Press2007}. Each fitted points is assigned initially a standard deviation calculated as the quadratic mean of the observational uncertainties. 

Even if our system is overdetermined ($n = 8$ equations for 5 parameters), the solution can be singular, or numerically very close to singular. In that case, we reduce the system to 3 parameters by fixing the coordinates of the conic center ($x_{0}, y_{0}$). Then we define a grid of values with a given step size, and we solve the equations for each point of the grid, searching for the conics which have $\Delta\chi^2 < 14.2$, the 3--$\sigma$ confidence region for the normal distribution with 3 degrees of freedom \citep{Press2007}.

Among the observers, two could have observed the occultation by Romulus, namely M. Bretton and V. Fristot (c.f. Tab. \ref{tbl-timings} and \ref{tbl-observers}). The first one made a naked eye observation and stop too early to observe Romulus event. The analysis of the video acquired by the second confirms that no occultation of Romulus occurred along his path. This result sets a strong upper limit on the size of Romulus, which allows us to reject all the fitted conics which intersect this chord.

\begin{figure}
  \includegraphics[width=\textwidth]{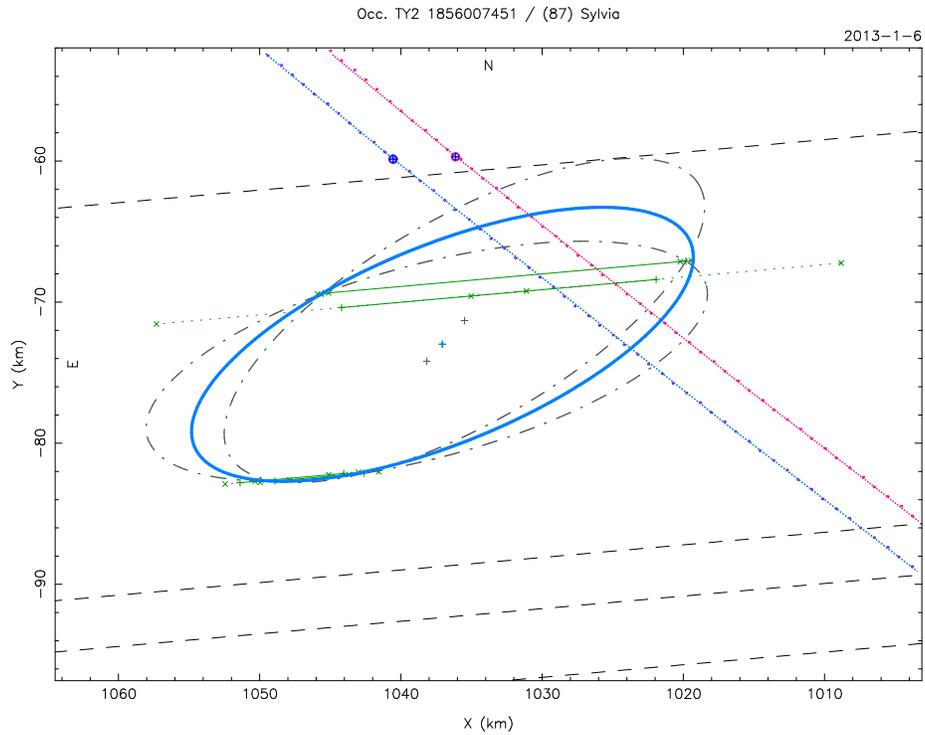}
  \caption{Romulus conic model fitted on the observed chords (green segments). The (blue) ellipse is the mean model, and the two dashed ellipses are the two extreme solutions at 3--$\sigma$. The dashed black lines are negative observations (northern one by V. Fristot). The blue (leftmost) and red lines show the predicted orbit of Romulus projected into the occultation plane, respectively based on the previous and the new orbital models. The two marks on these lines (located at $y \sim -60$ km) enlighten the predicted positions of Romulus at the epoch of occultation. \label{fig-romulusbestmodel}}
\end{figure}

Considering a range of 20 km wide for the coordinates of the conic center, and a grid step size of $10^{-3}$\,km, the best-fit conic profile of Romulus is (Fig. \ref{fig-romulusbestmodel}):
\[
 \begin{array}{lcrl}
   a    & = &  19.0\;\textrm{km}   & \pm\,1.6\,\textrm{km} \\
   b    & = &   7.0\;\textrm{km}   & \pm\,0.4\,\textrm{km} \\
   \psi & = & 112.4\,\arcdeg~~~    & \pm\,13.2\,\arcdeg \\
 \end{array}
\]
with uncertainties at 3--$\sigma$. It implies a mean axis ratio $a/b = 2.7 \pm\,0.3$, and a surface-area equivalent diameter $\deqS = 23.1 \pm\,0.7$ km.


\rev{This result relies on the reliability of the timings of the two northern chords of Romulus. Even if P. Tanga failed to report accurate absolute timings, he was able to estimate them using the computer clock (synchronized by NTP) used to control the camera, and to measure precisely the duration of events and the delay between the two events (see Tab. \ref{tbl-timings}). Using the 3-D shape model of Sylvia projected into the occultation plane, we can tie the time of the first disappearance of the star observed by P. Tanga to other timings with an accuracy smaller than 0.2s. In the case we shift Tanga's chords by +0.2s, the axis ratio of the best-fit profile of Romulus decreases to 2.2, with $a = 18$ km, $b = 8$ km, $\deqS = 24$ km, and equivalent uncertainties.}

\section{Dynamical overview on Sylvia system}\label{orbits}

Orbital solutions for Sylvia's satellites have been recently published by \cite{Fang2012}, and previously by \cite{Marchis2005a} at the occasion of the discovery of the second satellite. We take the opportunity of the successful observation of a stellar occultation by Romulus to test the reliability of its orbital solution, used to predict the event \citep{Berthier2013}, and to improve the orbital parameters of the two satellites. 

 \subsection{Reliability of Romulus orbit}
 
We used our genetic-based algorithm \emph{Genoid-Kepler} \citep{Vachier2012} to predict the position of Sylvia's satellites for the January 6, 2013 stellar occultation. The dynamics used to model the orbit of the satellites was a pure 2-body Keplerian problem (no mutual perturbation between the two satellites was applied). For Remus, we used 25 astrometric positions (acquired in 2001, 2004, 2010 and 2011) spanning 3948.7 days or 2910.6 revolutions. For Romulus, we used 65 astrometric positions (acquired in 2001, 2004, 2005, 2006, 2010 and 2011) spanning 3953.8 days or 1317.9 revolutions. Tables \ref{tbl-observingConditions}, \ref{tbl-obsRemus} and \ref{tbl-obsRomulus} detail the observing condition and the astrometric positions used for this work. The best solutions obtained give orbits with a RMS of 12.1 mas for Remus, and 11.2 mas for Romulus.

The successful observation of the occultation by Romulus highlights the reliability of our dynamical model. Figure
\ref{fig-romulusbestmodel} shows the goodness of the prediction of Romulus position with respect to Sylvia's center of mass (as defined by the 3-D model). The offset between the observed position of Romulus (e.g., defined by the center of the fitted profile) and its predicted position is $\Delta x = -3.3$ km, $\Delta y = -13.1$ km ($x$ and $y$ are positive toward East and North). This predicted position is located at 13.5 km from the observed position. In the occultation plane, the distance between the fitted profiles of Sylvia and Romulus is 1154.6 km. The geocentric orbital velocity of Sylvia being 17.1 km/s, the occultation by Romulus occurred 67.6 s later than Sylvia event. At this epoch, January 6, 2013 01:38:14.02 UTC, we infer the astrometric position of Romulus relatively to Sylvia's center of mass: $(x, y) = (590, -113) \pm\,1.5$ mas.


 \subsection{New orbital solutions}\label{newsoluce}

\begin{figure} 
  \includegraphics[width=\textwidth]{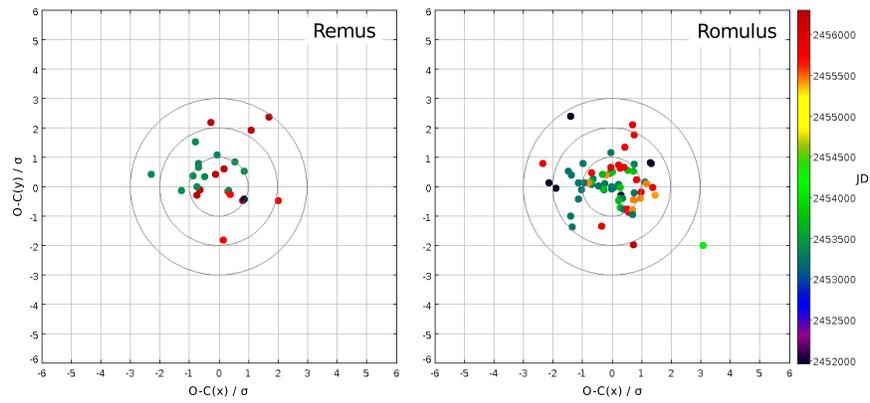}
  \caption{Mean residuals (observed minus computed satellite -- primary positions) normalized by the positional uncertainty $\sigma$ for Remus and Romulus.\label{fig-omc-sigma}}
\end{figure}

We added this astrometric position into \emph{Genoid-Kepler} input dataset, as well as an astrometric position acquired in 2010
with VLT/SINFONI \citep{Marchis2013a}. It increases the time span of data from 3953.8 to 4338.7 days or 1446.2 revolutions. This does not change significantly the dynamical solution of Romulus (RMS = 11.2 mas). Table \ref{tbl-dynamicalparam} presents the orbital parameters of Sylvia's satellites obtained with \emph{Genoid-Kepler}, and figure \ref{fig-omc-sigma} displays the residuals (the difference between the observed $O$ and computed $C$ positions of the satellite and the primary) normalized by the positional uncertainty $\sigma$. The orbital solutions indicate that the satellites follow a quasi circular orbit, nearly coplanar (very low mutual inclination, lesser than $0\arcdeg.5$), and nearly aligned with the primary equatorial plane (by $\sim$4\arcdeg). The average residuals are $dx = -0.4 \pm\,11.9$ mas, $dy = 4.7 \pm\,11.5$ mas for Remus, and $dx = 0.5 \pm\,12.7$ mas, $dy = 0.6 \pm\,9.4$ mas for Romulus (uncertainties at 3-$\sigma$), matching the level of accuracy of the observations. At the epoch of the occultation, the reliability of the model can be estimated by the difference between the observed and computed positions of Romulus, which are $dx = 0.2$ mas and $dy = -8.9$ mas, or $\Delta x = 1.0$ km and $\Delta y = -13.3$ km in the occultation plane. We inserted these orbital solutions in the Web service \href{http://vo.imcce.fr/webservices/miriade}{Miriade}\footnote{http://vo.imcce.fr/webservices/miriade/} of our Virtual Observatory Solar system portal, allowing everyone to compute the ephemeris of Sylvia's satellites.


The mass of Sylvia can be derived from the best-fit keplerian orbit of the satellites (see Tab. \ref{tbl-dynamicalparam}). Assuming the Gaussian gravitational constant and the solar mass, one can easily derive, independently, the mass of Sylvia from the semi-major axis and the orbital period of each satellite (neglecting its own mass). We find $M_{S} = 1.380\,\pm 0.151 \; 10^{19}$ kg from Remus, and $M_{S} = 1.476\,\pm 0.128\; 10^{19}$ kg from Romulus, considering the 1--$\sigma$ uncertainties of the parameters. It is interesting to note that the two masses are consistent within their uncertainties. The mass derived from Romulus orbit is also very similar, by $\sim$0.5\%, to that found by \cite{Fang2012}, although it has been obtained by the latter by the fit of a fully dynamical three-body model.

From the mass of Sylvia and its volume, defined by its 3-D shape model, we infer its density. We find 1.29 and 1.38 g.cm$^{-3}$ from Remus and Romulus data respectively, with an uncertainty of $\,\pm 0.16$ g.cm$^{-3}$, taking into account the 1--$\sigma$ uncertainties of the parameters. The resulting density of 1.34\,$\pm$\,0.21 g.cm$^{-3}$ is slightly lower, although consistent, with the average density of X-type asteroids of 1.85\,$\pm$\,0.81 g.cm$^{-3}$ \citep{Carry2012c}. The composition and meteorite analogs of X-types is still debated, and proposed analogs encompass CV and CO carbonaceous chondrites, enstatite chondrites, aubrites, mesosiderites, and even iron, stony-iron meteorites \citep{2005-AA-430-Barucci, 2009-Icarus-202-Vernazza, 2011-Icarus-216-Vernazza, 2010-Icarus-210-Ockert-Bell, 2011-Icarus-214-Fornasier}, which density ranges from 2.8 to 7.7 g.cm$^{-3}$. Although we do not have information on the composition of Sylvia, its density, lower than that of its most-likely components, indicates the presence of voids and/or pockets of very low density material (e.g., ices). The macroporosity we derive here ranges from 52\% to 82\% depending on the density of the analog meteorite.

 \subsection{Examination of the quadrupole term $J_{2}$}\label{examJ2}

\begin{figure} 
  \includegraphics[width=\textwidth]{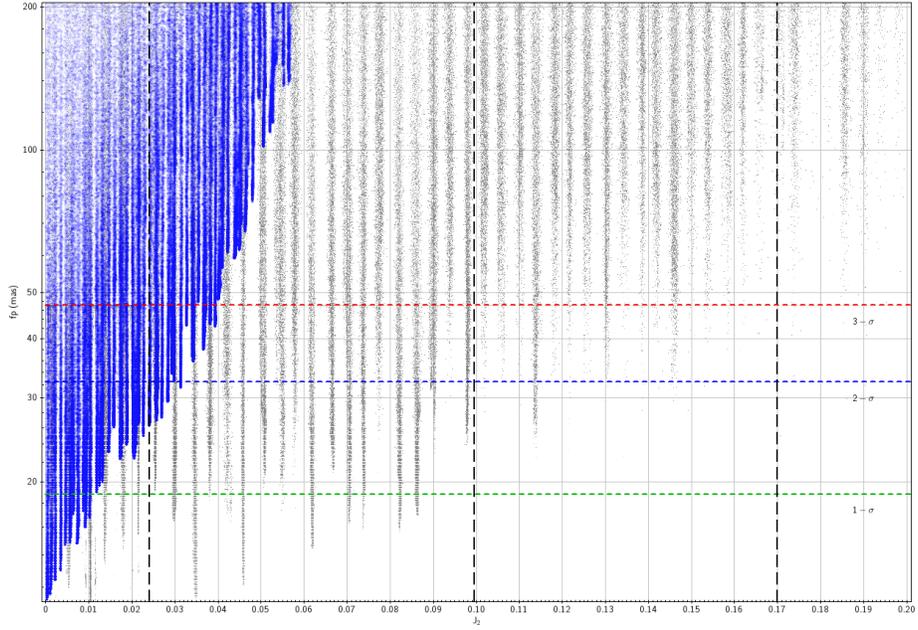}
  \caption{Result of the examination of the quadrupole term $J_2$ with \emph{Genoid-ANIS}. The gray dots are all solutions. The (blue) biggest points show the solutions for which the orbital pole covers a range of $20\arcdeg$ around the best-fit spin solution of the primary (i.e., rectangles in Fig. \ref{fig-rmspole}). The (green, blue, red) horizontal dashed lines correspond, respectively, to 1--$\sigma$, 2--$\sigma$ and 3--$\sigma$ thresholds. The three vertical lines indicate the $J_2$ values estimated, from left to right, from the 3-D shape model (see Sec. \ref{sylvia}), by \cite{Fang2012} and by \cite{Marchis2005a}. The best candidate solutions are obtained for $J_2 \rightarrow 0$.\label{fig-J2}}
\end{figure}

We used our \emph{Genoid-ANIS} algorithm \citep{Vachier2012} to search for a possible influence of an irregular mass distribution, considering the amount of voids determined above. We initialized the problem with $J_2 = 0.024$, the value estimated from the 3-D shape model (see Sec. \ref{sylvia}). We search for orbital solutions in the space $0 \leq J_2 \leq 0.2$, all other parameters being free. The results are shown in Fig. \ref{fig-J2}. Assuming that the directions of the orbital poles of the satellites must be nearly aligned with the primary pole of rotation (hypothesis justified by the low inclination of orbital planes, see Sec. \ref{newsoluce}), we discard all solutions for which coordinates of orbital poles are outside the 3--$\sigma$ confidence interval of the primary spin-vector coordinates (see Sec. \ref{sylvia} and Fig. \ref{fig-rmspole}). We find that the best candidate solutions are obtained for $J_2 \rightarrow 0$. It means that no significant precession effect of the apsidal and nodal nodes is detected, and that the purely Keplerian orbit is enough to fit the available data to their accuracy.


\cite{Winter2009} showed that Sylvia's system is not stable unless the primary has a minimal amount of oblateness, at least 0.1\% of the assumed primary $J_2$ of 0.17 reported by \cite{Marchis2005a}. This oblateness provides a faster stabilizing effect on the satellites' orbits than other gravitational perturbations (e.g., Sun, Jupiter, see also \cite{Frouard2012}). Our $J_2 \sim 0$ result does not contradict this statement, insofar as we analyze observational data and do not study the long term stability of the system. It shows that the estimation of the quadrupole term $J_{2}$ of the gravitational potential of Sylvia is not obvious, even impossible, given the available observational data, especially since a keplerian model provides the best-fit orbital solution. The discrepancy between our result and those of \cite{Fang2012} (who found a primary $J_2$ of 0.1), or those of \cite{Marchis2005a}, reveals that the estimation of the $J_{2}$ term is strongly correlated with the orbit-fitting method. In all cases, very similar orbital solutions are obtained, and the same conclusion applies: Sylvia's system is ancient and in a very stable state.

\section{Discussion}

 \subsection{Interior of (87) Sylvia}

The difference in $J_2$ from our dynamical analysis and as derived from the 3-D shape model implies that the assumption of an homogeneous mass distribution is not valid. The $J_2 \sim 0$ determined dynamically indicates a more concentrated mass distribution than the 3-D shape suggests, so that the primary could be differentiated with a dense core. The same result emerged from the analysis of the orbit of the 28 km-diameter satellite of (22) Kalliope, a 166 km-diameter M-type asteroid. \citet{Vachier2012} suggested a differentiated internal structure for the primary of this binary system.

Recent developments \citep[e.g.,][]{Ghosh2003co} in thermal modeling of small solar system bodies showed that large enough (radius $>$ 7 km) asteroids that accreted shortly after the Calcium-Aluminum inclusion formation ($<$ 2.5 Myr), when $^{26}$Al was still abundant, could have been molten internally. As they were cooling off, their internal composition became partially differentiated. Those models showed that these asteroids remained covered with a thick (up to tens of kilometers) un-melted relict crust \citep[see a review on this mechanism and a thorough discussion in][]{Weiss2013ax}.

Although we cannot conclude whether or not the internal structure of (87) Sylvia is differentiated, the very low $J_2$, perturbing the satellite's orbit, hints toward a dense core embedded into an irregularly shaped material. The lower density of the surrounding material may be related to composition or structure, e.g., a large macro-porosity generated by fractures.

 \subsection{Elongated shape of Romulus}

The shape and size of Romulus was estimated in section \ref{romulus} from the chords of an occultation, leading to a surface-area equivalent diameter $\deqS = 23.1 \pm 0.7$ km (H = 12.1--12.3, assuming the same albedo as the primary of 0.0435, \citealt{Tedesco2002}), and an axis ratio $a/b \sim 2.7 \pm 0.3$ (assuming an ellipsoid). This is the second time the shape and size of a satellite of asteroid is measured using this technique. An occultation by Linus, satellite of the asteroid (22) Kalliope, gave a similar opportunity \citep{Descamps2008a}, but the chords were too close to each other and the shape could only be approximated by a circular fit. In the case of Romulus, the positive and negative chords are sufficiently spread (Fig. \ref{fig-romulusbestmodel}) to give a meaningful constraint on its shape. The satellite appears extremely elongated, with the main axis of the silhouette ellipse oriented at $12\arcdeg \pm 4\arcdeg$ from the primary. 

\begin{figure} 
  \includegraphics[width=\textwidth]{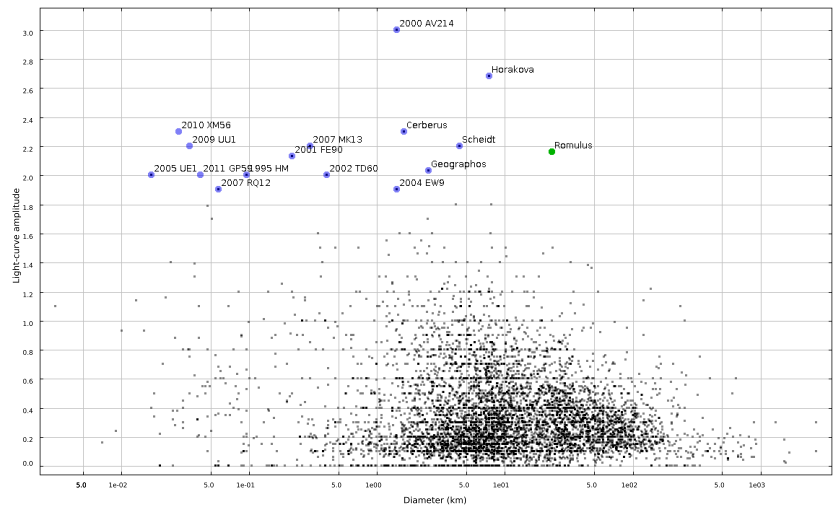}
  \caption{Distribution of 6160 asteroids based on light-curve amplitude versus diameter (source of data: \emph{Asteroid Lightcurve Database}, \citealp{Warner2009d}). High amplitude objects are labeled. We added Romulus for comparison, with $\deqS = 23.1$ km and an amplitude estimated to 2.16 (corresponding to an axis ratio of 2.7).\label{fig-lcdb}}
\end{figure}

We searched in the \emph{Asteroid Lightcurve Database} \citep{Warner2009d} containing the light-curves of 6160 small solar system bodies (including NEAs, TNOs, main-belt and a few comets) those with an amplitude greater than 1.9 (corresponding to a size ratio $\ge$ 2.4) and found 15 asteroids, including 12 NEAs and 3 main-belt asteroids (Fig. \ref{fig-lcdb}). They are all significantly smaller than Romulus with a diameter varying from 20 m to 7.5 km. The largest, (44530) Horakova, has a light-curve poorly constrained with a period $P = 160$ h and a maximum light-curve amplitude $\Delta m \sim 2.7$, suggesting that it could be a tidally locked binary. Asteroid (1620) Geographos ($D = 1.6$ km, $\Delta m \sim 2.0$) is smaller but shows the same light-curve amplitude as Romulus. Moreover its shape is well-defined thanks to delay-Doppler radar observations \citep{Ostro1996b}. Shape reconstruction from radar gives $a/b = 2.76 \pm 0.18$, similar to Romulus. \cite{Bottke1999a} has shown by numerical simulations that tidal disruptions during a close encounter with a planet could produce the elongated shape of Geographos. Similarly, because Romulus is relatively close to the primary ($\sim$10 $\times$ Sylvia's radius), and could be a rubble-pile satellite which formed from fragments of a catastrophic collision on the parent body, its elongated shape may result of the tidal forces from the elongated and spinning primary. Another possibility is that Romulus is a bilobed satellite, relic of its formation by a low-relative speed encounter of two $\sim$10 km fragments of the parent body and captured in the gravitational field of the primary. An accurate modeling of the tidal evolution of the orbit and shape of the satellite could shed light on the internal stress and cohesive forces of the satellite.


\section{Conclusion}

This work demonstrates once again that the combination of adaptive-optics observations with light-curve photometric observations and stellar occultations is a powerful way to study multiple asteroid system. Similarly to our work on the triple asteroid (93) Minerva \citep{Marchis2013}, we derive here the 3-D shape and size of Sylvia's primary and constrain its internal structure. The successful observation of the occultation by Romulus, the outer satellite of the system, provides the first well-constrained estimate of the shape of an asteroid's moon which has an extreme elongation, likely due to the tides due to the primary. A follow-up of the orbits of Sylvia's moons based on additional AO observations, recorded with present AO systems and on the next generation of adaptive optics currently being designed for the Keck telescope \citep{Wizinowich2010} or set up on new telescopes (e.g., LBTAO, see \citealt{Esposito2011}), will help to confirm the absence of precession indicative of a heterogeneous interior for this asteroid. 

\begin{figure}
  \includegraphics[width=\textwidth]{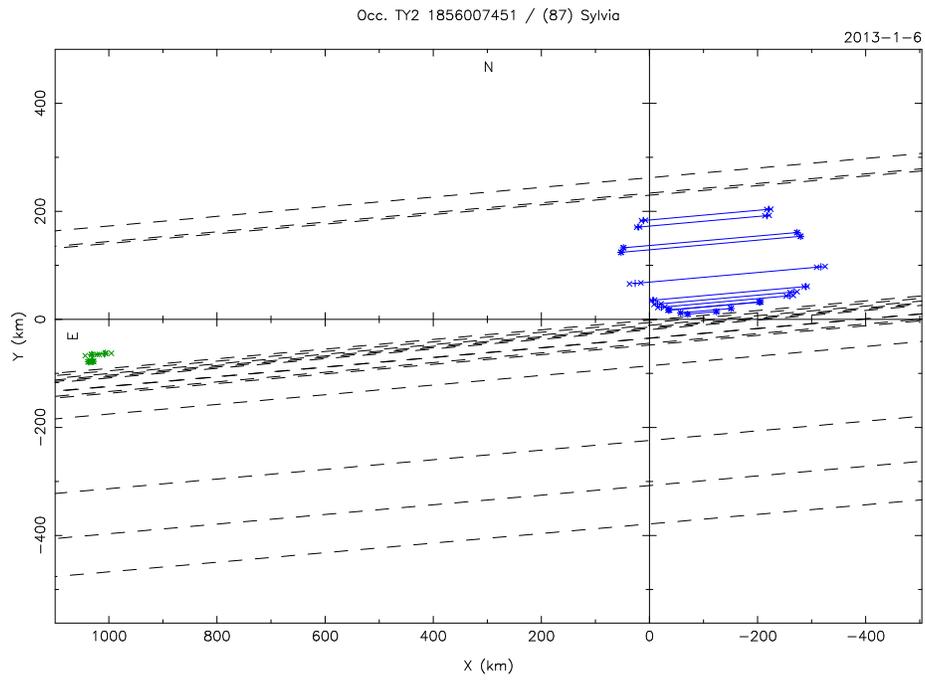}
  \caption{Observed chords (blue segments) of the January 6, 2013 occultation of TYC2 1856-00745-1 by (87) Sylvia and its satellite Romulus (green leftmost segments). The black dashed lines represent negative observations. \label{fig-occhords}}
\end{figure}

The analysis of the occultation chords recorded on January 6, 2013 does not reveal the presence of Remus, the inner satellite, because of a poor coverage of the Northern part of the occultation path (Fig. \ref{fig-occhords}), where Remus was predicted to be located. No observer reported occultation events due to the presence of smaller, yet unknown, satellites around Sylvia's primary. From the timing accuracy of recorded light-curves (Tab. \ref{tbl-timings}), we estimate that observers could have detected other satellites around Sylvia with a diameter larger than 2 km along the occultation chords (Fig.~\ref{fig-occhords}). For comparison, the upper limit of detection for the AO observations varies with the distance to the primary and the quality of the AO correction; typically a satellite larger than $\sim$5 km in diameter could have been detected if located at Remus' distance ($\sim0.3\arcsec$). This is a clear illustration of the usefulness of occultation that help determining the multiplicity of an asteroid, even if it is far, too faint, and/or accompanied with a small moon undetectable by AO systems. As an example, the observation of an occultation by the Jupiter-trojan asteroid (911) Agamemnon in 2012 showed a deep brief secondary event that is likely due to a satellite of about 5 km \citep{Timerson2013}. Additional occultations involving (87) Sylvia are predicted in the near future. Table \ref{tbl-futurpred} lists a selection of events scheduled for the next 10 years, and their locations. In the future, we will continue to monitor the orbits of the satellites and deliver predictions of the moons' paths at the time of occultations. 


\bigskip
Acknowledgments
\medskip

We are grateful to all the observers for their efforts and for providing their observations (Tab. \ref{tbl-observers}). We appreciate the help of S. Preston (IOTA) who made available the prediction of the event, to H. Harris (UNSO/Flagstaff) and B. Owen (TMO) for providing last minute astrometric observations of the asteroid, to P. Descamps and {\em Les Makes} observatory (La R\'eunion island) for the astrometric measurements of the occulted star, to E. Frappa (Euraster), J. Lecacheux (LESIA/Observatoire de Paris) et al. to have organized the observational campaign and collected the observer reports. The work of F. Marchis was supported by the NASA grant NNX11AD62G. The work of J. {\v D}urech was supported by the grants GACR 209/10/0537 and P209/12/0229 of the Czech Science Foundation. We thank the C2PU team who gave us access to the data collected by the {\em Omicron} telescope of the {\em Centre P\'edagogique Plan\`ete et Univers} (Observatoire de la C\^ote d'Azur, Calern site), as well as the TAROT observatory team who made available their robotic telescopes to record occultation events.

\clearpage

\begin{small}
\begin{longtable}{ccccccccc}
 \caption{Observational conditions of astrometric positions used for this work. References: a, b, c: reanalysis of data collected respectively by \cite{Brown2001}; \cite{Storrs2001}; \cite{Marchis2005a}; d: astrometric positions reported in \cite{Fang2012}; e, f: unpublished data acquired respectively by Marchis et al. and Carry et al.; g: this work} \label{tbl-observingConditions} \\

\hline \noalign{\smallskip}
Date time &   V   & D$_{Earth}$ & D$_{Sun}$ & Phase     & Elong.     & AM & Telescope & Ref. \\
(UTC)     & (mag) & (au)        & (au)      & (\arcdeg) & (\arcdeg)  &    &           &      \\
\hline \noalign{\smallskip}
\endfirsthead

\multicolumn{9}{c}%
{{\bfseries \tablename\ \thetable{} -- continued}} \\

\hline \noalign{\smallskip}
Date time &   V   & D$_{Earth}$ & D$_{Sun}$ & Phase     & Elong.     & AM & Telescope & Ref. \\
(UTC)     & (mag) & (au)        & (au)      & (\arcdeg) & (\arcdeg)  &    &           &      \\
\hline \noalign{\smallskip}
\endhead

\hline \noalign{\smallskip}
\multicolumn{9}{r}{{Continued on next page}} \\
\endfoot

\noalign{\smallskip}\hline
\endlastfoot
2001-02-19 08:39:46.94 & 12.41 & 2.791595 & 3.755543 &  3.95 & 164.84 & 1.123 & Keck/NIRC2   & a \\
2001-02-19 10:03:57.89 & 12.41 & 2.791677 & 3.755554 &  3.96 & 164.81 & 1.014 & Keck/NIRC2   & a \\
2001-02-19 10:12:43.20 & 12.41 & 2.791686 & 3.755555 &  3.96 & 164.81 & 1.010 & Keck/NIRC2   & a \\
2001-02-20 09:39:27.36 & 12.41 & 2.793272 & 3.755729 &  4.08 & 164.33 & 1.027 & Keck/NIRC2   & a \\
2001-02-20 09:46:55.78 & 12.41 & 2.793281 & 3.755730 &  4.08 & 164.33 & 1.021 & Keck/NIRC2   & a \\
2001-02-23 13:40:12.58 & 12.45 & 2.800411 & 3.756284 &  4.58 & 162.37 &  -    & HST/WFPC2    & b \\[0.2cm]

2004-07-25 10:22:17.76 & 12.14 & 2.410849 & 3.218312 & 12.82 & 135.39 & 1.205 & VLT/NACO     & c \\
2004-08-10 07:17:16.22 & 11.89 & 2.293127 & 3.217288 &  8.99 & 150.30 & 1.008 & VLT/NACO     & c \\
2004-08-29 07:18:19.30 & 11.64 & 2.230743 & 3.216817 &  5.05 & 163.73 & 1.115 & VLT/NACO     & c \\
2004-08-29 07:26:17.95 & 11.64 & 2.230739 & 3.216817 &  5.05 & 163.73 & 1.133 & VLT/NACO     & c \\
2004-08-29 08:46:08.83 & 11.64 & 2.230705 & 3.216817 &  5.04 & 163.75 & 1.445 & VLT/NACO     & c \\
2004-09-01 05:54:55.01 & 11.63 & 2.229630 & 3.216818 &  4.89 & 164.26 & 1.017 & VLT/NACO     & c \\
2004-09-01 06:03:55.01 & 11.63 & 2.229631 & 3.216818 &  4.89 & 164.26 & 1.024 & VLT/NACO     & c \\
2004-09-01 08:25:31.58 & 11.63 & 2.229646 & 3.216818 &  4.89 & 164.27 & 1.406 & VLT/NACO     & c \\
2004-09-03 07:09:06.34 & 11.63 & 2.230255 & 3.216830 &  4.90 & 164.22 & 1.148 & VLT/NACO     & c \\
2004-09-04 08:41:35.81 & 11.64 & 2.231039 & 3.216840 &  4.94 & 164.07 & 1.601 & VLT/NACO     & c \\
2004-09-05 04:09:27.94 & 11.64 & 2.231820 & 3.216849 &  4.99 & 163.89 & 1.015 & VLT/NACO     & c \\
2004-09-05 08:15:20.74 & 11.64 & 2.232024 & 3.216852 &  5.01 & 163.85 & 1.452 & VLT/NACO     & c \\
2004-09-06 03:50:04.13 & 11.65 & 2.233033 & 3.216863 &  5.08 & 163.61 & 1.027 & VLT/NACO     & c \\
2004-09-07 02:36:23.04 & 11.65 & 2.234453 & 3.216878 &  5.18 & 163.28 & 1.142 & VLT/NACO     & c \\
2004-09-07 09:24:57.31 & 11.65 & 2.234943 & 3.216883 &  5.21 & 163.17 & 2.283 & VLT/NACO     & c \\
2004-09-08 06:54:52.01 & 11.66 & 2.236557 & 3.216899 &  5.33 & 162.78 & 1.171 & VLT/NACO     & c \\
2004-09-11 04:51:29.95 & 11.69 & 2.243369 & 3.216966 &  5.80 & 161.20 & 1.006 & VLT/NACO     & c \\
2004-09-13 03:25:24.10 & 11.72 & 2.249181 & 3.217022 &  6.17 & 159.92 & 1.020 & VLT/NACO     & c \\
2004-09-13 05:30:10.66 & 11.72 & 2.249465 & 3.217024 &  6.19 & 159.86 & 1.047 & VLT/NACO     & c \\
2004-09-14 03:49:52.03 & 11.73 & 2.252630 & 3.217054 &  6.39 & 159.19 & 1.003 & VLT/NACO     & c \\
2004-09-14 06:36:31.10 & 11.73 & 2.253047 & 3.217058 &  6.41 & 159.10 & 1.199 & VLT/NACO     & c \\
2004-09-14 06:44:34.08 & 11.73 & 2.253068 & 3.217058 &  6.41 & 159.10 & 1.225 & VLT/NACO     & c \\
2004-09-15 05:08:21.70 & 11.75 & 2.256499 & 3.217090 &  6.62 & 158.39 & 1.032 & VLT/NACO     & c \\
2004-09-15 05:16:22.94 & 11.75 & 2.256520 & 3.217090 &  6.62 & 158.39 & 1.042 & VLT/NACO     & c \\
2004-10-19 00:27:06.91 & 12.30 & 2.518403 & 3.219585 & 14.37 & 126.79 & 1.038 & VLT/NACO     & c \\
2004-10-20 00:08:28.03 & 12.31 & 2.529405 & 3.219697 & 14.54 & 125.85 & 1.057 & VLT/NACO     & c \\
2004-10-25 06:28:30.72 & 12.39 & 2.590441 & 3.220330 & 15.39 & 120.89 & 1.359 & Keck/NIRC2   & c \\
2004-11-02 07:34:37.34 & 12.51 & 2.690426 & 3.221417 & 16.43 & 113.48 & 1.428 & Gemini/NIRI  & c \\[0.2cm]

2005-11-01 13:18:57.31 & 11.99 & 2.442235 & 3.394667 &  5.81 & 159.75 & 1.119 & Gemini/NIRI  & e \\
2005-11-06 08:19:53.76 & 11.90 & 2.426453 & 3.398045 &  4.28 & 165.18 & 1.304 & Gemini/NIRI  & e \\
2005-11-06 08:31:43.10 & 11.90 & 2.426431 & 3.398051 &  4.28 & 165.19 & 1.254 & Gemini/NIRI  & e \\
2005-12-20 10:01:06.82 & 12.35 & 2.598025 & 3.429626 & 10.23 & 141.83 & 1.201 & Gemini/NIRI  & e \\
2005-12-20 10:08:25.73 & 12.35 & 2.598075 & 3.429630 & 10.23 & 141.82 & 1.226 & Gemini/NIRI  & e \\
2005-12-21 08:42:06.05 & 12.36 & 2.607454 & 3.430311 & 10.46 & 140.78 & 1.041 & Gemini/NIRI  & e \\
2005-12-21 08:47:19.68 & 12.36 & 2.607491 & 3.430314 & 10.46 & 140.78 & 1.048 & Gemini/NIRI  & e \\[0.2cm]

2006-01-01 10:43:07.10 & 12.56 & 2.731924 & 3.438370 & 12.90 & 128.79 & 1.765 & Gemini/NIRI  & e \\
2006-01-06 09:14:36.10 & 12.65 & 2.794520 & 3.441971 & 13.78 & 123.64 & 1.295 & Gemini/NIRI  & e \\
2006-01-06 09:20:56.26 & 12.65 & 2.794578 & 3.441974 & 13.78 & 123.64 & 1.323 & Gemini/NIRI  & e \\
2006-12-12 16:07:33.89 & 12.73 & 2.883229 & 3.670720 & 10.49 & 137.29 & 1.267 & Keck/NIRC2   & e \\[0.2cm]

2010-07-27 05:23:00.04 & 11.64 & 2.259758 & 3.253753 &  4.03 & 166.98 & 1.027 & VLT/SINFONI  & e \\
2010-08-15 08:37:41.66 & 11.94 & 2.331035 & 3.244915 &  8.95 & 150.09 & 1.639 & Gemini/NIRI  & e \\
2010-08-25 08:28:46.85 & 12.10 & 2.403504 & 3.240567 & 11.54 & 140.07 & 1.662 & Gemini/NIRI  & e \\
2010-08-28 08:22:10.27 & 12.15 & 2.429262 & 3.239301 & 12.25 & 137.09 & 1.671 & Gemini/NIRI  & e \\
2010-08-30 05:57:29.00 & 12.18 & 2.446480 & 3.238507 & 12.68 & 135.21 & 1.647 & VLT/NCAO     & f \\
2010-09-01 08:27:56.74 & 12.22 & 2.466323 & 3.237635 & 13.13 & 133.14 & 1.717 & Gemini/NIRI  & e \\
2010-09-02 06:34:03.36 & 12.23 & 2.475252 & 3.237257 & 13.33 & 132.24 & 1.694 & Gemini/NIRI  & e \\[0.2cm]

2011-10-07 02:37:58.08 & 11.68 & 2.277797 & 3.251690 &  4.40 & 165.53 & 1.357 & VLT/NACO     & d \\
2011-11-06 02:03:15.84 & 11.99 & 2.376961 & 3.266863 &  8.84 & 149.55 & 1.088 & VLT/NACO     & d \\
2011-11-08 03:05:36.96 & 12.03 & 2.392516 & 3.267956 &  9.38 & 147.45 & 1.066 & VLT/NACO     & d \\
2011-11-10 01:04:56.64 & 12.06 & 2.408016 & 3.268988 &  9.88 & 145.48 & 1.149 & VLT/NACO     & d \\
2011-11-15 00:36:51.84 & 12.15 & 2.452228 & 3.271698 & 11.12 & 140.34 & 1.163 & VLT/NACO     & d \\
2011-11-16 01:07:06.24 & 12.17 & 2.461963 & 3.272259 & 11.36 & 139.28 & 1.103 & VLT/NACO     & d \\
2011-11-20 01:06:14.40 & 12.24 & 2.502190 & 3.274473 & 12.27 & 135.18 & 1.086 & VLT/NACO     & d \\
2011-12-15 05:06:34.56 & 12.65 & 2.816075 & 3.289009 & 16.26 & 110.50 & 1.104 & Keck/NIRC2   & d \\
2011-12-15 05:29:28.32 & 12.65 & 2.816297 & 3.289018 & 16.27 & 110.48 & 1.087 & Keck/NIRC2   & d \\
2011-12-15 06:28:30.72 & 12.65 & 2.816873 & 3.289043 & 16.27 & 110.44 & 1.098 & Keck/NIRC2   & d \\
2011-12-16 04:30:17.28 & 12.66 & 2.829790 & 3.289592 & 16.36 & 109.59 & 1.147 & Keck/NIRC2   & d \\
2011-12-16 06:01:26.40 & 12.67 & 2.830682 & 3.289630 & 16.36 & 109.53 & 1.084 & Keck/NIRC2   & d \\
2011-12-17 06:16:33.60 & 12.68 & 2.844990 & 3.290237 & 16.45 & 108.60 & 1.093 & Keck/NIRC2   & d \\
2011-12-18 04:43:58.08 & 12.69 & 2.858315 & 3.290799 & 16.53 & 107.74 & 1.114 & Keck/NIRC2   & d \\
2011-12-18 04:53:02.40 & 12.69 & 2.858404 & 3.290803 & 16.53 & 107.73 & 1.104 & Keck/NIRC2   & d \\[0.2cm]

2013-01-06 01:38:14.18 & 12.30 & 2.654901 & 3.581037 &  6.01 & 157.56 & 1.438 & Occ. data    & g \\

\end{longtable}
\end{small}

\clearpage

\begin{small}
\begin{longtable}{crrrrrrcc}
 \caption{List of \textbf{Remus} astrometric observations ($X_o$, $Y_o$) collected from February 2001 to December 2011. The computed positions ($X_c$, $Y_c$) were obtained with the orbital elements  published in this paper. The  uncertainty $\sigma$ of astrometric positions is taken as the instrument plate scale. References: c.f. Tab. \ref{tbl-observingConditions}.}  \label{tbl-obsRemus} \\

\hline \noalign{\smallskip}
Date time & \multicolumn{1}{c}{$X_o$} & \multicolumn{1}{c}{ $Y_o$} &  \multicolumn{1}{c}{$X_c$} &  \multicolumn{1}{c}{$Y_c$} &  \multicolumn{1}{c}{$X_{O-C}$} &  \multicolumn{1}{c}{$Y_{O-C}$} & $\sigma$ & Ref. \\
(UTC) & (arcsec) & (arcsec) & (arcsec) & (arcsec) & (arcsec) & (arcsec) & (mas) & \\
\hline \noalign{\smallskip}
\endfirsthead

\multicolumn{9}{c}%
{{\bfseries \tablename\ \thetable{} -- continued}} \\

\hline \noalign{\smallskip}
Date time & \multicolumn{1}{c}{$X_o$} & \multicolumn{1}{c}{ $Y_o$} &  \multicolumn{1}{c}{$X_c$} &  \multicolumn{1}{c}{$Y_c$} &  \multicolumn{1}{c}{$X_{O-C}$} &  \multicolumn{1}{c}{$Y_{O-C}$} & $\sigma$ & Ref. \\
(UTC) & (arcsec) & (arcsec) & (arcsec) & (arcsec) & (arcsec) & (arcsec) & (mas) & \\
\hline \noalign{\smallskip}
\endhead

\hline \noalign{\smallskip}
\multicolumn{9}{r}{{Continued on next page}} \\
\endfoot

\noalign{\smallskip}\hline
\endlastfoot

2001-02-23T13:40:12.58 &  0.3448 &  0.0290 &  0.3347 &  0.0341 &  0.0101 & -0.0051 & 40.2 & b \\[0.2cm]

2004-08-10T07:19:16.00 &  0.3952 &  0.0136 &  0.4046 &  0.0032 & -0.0094 &  0.0104 & 13.3 & c \\
2004-09-01T05:57:01.00 &  0.2074 & -0.0881 &  0.2379 & -0.0936 & -0.0306 &  0.0054 & 13.3 & c \\
2004-09-01T06:05:55.00 &  0.2216 & -0.0904 &  0.2278 & -0.0948 & -0.0063 &  0.0044 & 13.3 & c \\
2004-09-03T07:11:06.00 & -0.2199 &  0.1030 & -0.2082 &  0.0981 & -0.0117 &  0.0049 & 13.3 & c \\
2004-09-05T04:11:28.00 &  0.4052 & -0.0507 &  0.4007 & -0.0489 &  0.0045 & -0.0018 & 13.3 & c \\
2004-09-07T02:38:22.00 & -0.4350 &  0.0009 & -0.4182 &  0.0029 & -0.0168 & -0.0020 & 13.3 & c \\
2004-09-08T06:56:51.00 & -0.2461 & -0.0620 & -0.2451 & -0.0760 & -0.0010 &  0.0141 & 13.3 & c \\
2004-09-13T03:27:24.00 &  0.3693 &  0.0385 &  0.3786 &  0.0298 & -0.0093 &  0.0087 & 13.3 & c \\
2004-09-13T05:32:11.00 &  0.4086 & -0.0136 &  0.4187 & -0.0135 & -0.0101 & -0.0001 & 13.3 & c \\
2004-09-14T03:51:51.00 & -0.1897 &  0.1251 & -0.1793 &  0.1050 & -0.0105 &  0.0202 & 13.3 & c \\
2004-09-15T05:10:21.96 & -0.3956 &  0.0002 & -0.4028 & -0.0109 &  0.0073 &  0.0110 & 13.3 & c \\
2004-09-15T05:18:22.00 & -0.3944 & -0.0012 & -0.4055 & -0.0081 &  0.0112 &  0.0069 & 13.3 & c \\[0.2cm]

2010-08-15T08:37:52.00 &  0.2289 &  0.1330 &  0.2249 &  0.1351 &  0.0040 & -0.0021 & 13.3 & e \\
2010-08-28T08:23:47.00 & -0.3366 & -0.0499 & -0.3417 & -0.0461 &  0.0052 & -0.0038 & 13.3 & e \\
2010-09-01T08:29:10.00 & -0.2675 & -0.1232 & -0.2695 & -0.0989 &  0.0020 & -0.0243 & 13.3 & e \\
2010-09-02T06:35:15.00 &  0.3820 & -0.1122 &  0.3555 & -0.1057 &  0.0265 & -0.0066 & 13.3 & e \\[0.2cm]

2011-10-07T02:37:58.08 & -0.3910 & -0.0340 & -0.4017 & -0.0277 &  0.0107 & -0.0063 & 13.3 & d \\
2011-11-08T03:05:36.96 &  0.2180 &  0.0570 &  0.2218 &  0.0283 & -0.0038 &  0.0287 & 13.3 & d \\
2011-11-10T01:04:56.64 & -0.3420 &  0.0020 & -0.3564 & -0.0233 &  0.0144 &  0.0253 & 13.3 & d \\
2011-11-16T01:07:06.24 &  0.3810 &  0.0180 &  0.3827 &  0.0126 & -0.0017 &  0.0054 & 13.3 & d \\
2011-12-15T05:06:34.56 & -0.3410 & -0.0110 & -0.3346 & -0.0097 & -0.0064 & -0.0013 &  9.9 & d \\
2011-12-15T05:29:28.32 & -0.3420 & -0.0150 & -0.3347 & -0.0121 & -0.0073 & -0.0029 &  9.9 & d \\
2011-12-15T06:28:30.72 & -0.3250 & -0.0120 & -0.3266 & -0.0179 &  0.0016 &  0.0059 &  9.9 & d \\
2011-12-17T06:16:33.60 &  0.3480 &  0.0350 &  0.3314 &  0.0116 &  0.0166 &  0.0234 &  9.9 & d \\

\end{longtable}
\end{small}

\clearpage

\begin{small}
\begin{longtable}{crrrrrrcc}
 \caption{List of \textbf{Romulus} astrometric observations ($X_o$, $Y_o$) collected from February 2001 to December 2011. The computed positions ($X_c$, $Y_c$) were obtain with the orbital elements published in this paper. The  uncertainty $\sigma$ of astrometric positions is taken as the instrument plate scale. References: c.f. Tab. \ref{tbl-observingConditions}.}  \label{tbl-obsRomulus} \\

\hline \noalign{\smallskip}
Date time & \multicolumn{1}{c}{$X_o$} & \multicolumn{1}{c}{ $Y_o$} &  \multicolumn{1}{c}{$X_c$} &  \multicolumn{1}{c}{$Y_c$} &  \multicolumn{1}{c}{$X_{O-C}$} &  \multicolumn{1}{c}{$Y_{O-C}$} & $\sigma$ & Ref. \\
(UTC) & (arcsec) & (arcsec) & (arcsec) & (arcsec) & (arcsec) & (arcsec) & (mas) & \\
\hline \noalign{\smallskip}
\endfirsthead

\multicolumn{9}{c}%
{{\bfseries \tablename\ \thetable{} -- continued}} \\

\hline \noalign{\smallskip}
Date time & \multicolumn{1}{c}{$X_o$} & \multicolumn{1}{c}{ $Y_o$} &  \multicolumn{1}{c}{$X_c$} &  \multicolumn{1}{c}{$Y_c$} &  \multicolumn{1}{c}{$X_{O-C}$} &  \multicolumn{1}{c}{$Y_{O-C}$} & $\sigma$ & Ref. \\
(UTC) & (arcsec) & (arcsec) & (arcsec) & (arcsec) & (arcsec) & (arcsec) & (mas) & \\
\hline \noalign{\smallskip}
\endhead

\hline \noalign{\smallskip}
\multicolumn{9}{r}{{Continued on next page}} \\
\endfoot

\noalign{\smallskip}\hline
\endlastfoot

2001-02-19T08:41:16.998	&  0.3401 &  0.2101 &  0.3589 &  0.2108 & -0.0189 & -0.0007 &  9.9 & a \\
2001-02-19T10:04:18.001	&  0.2806 &  0.2190 &  0.3017 &  0.2178 & -0.0211 &  0.0012 &  9.9 & a \\
2001-02-19T10:13:03.002	&  0.2815 &  0.2421 &  0.2954 &  0.2185 & -0.0139 &  0.0236 &  9.9 & a \\
2001-02-20T09:40:55.997	& -0.6242 &  0.0306 & -0.6273 &  0.0337 &  0.0031 & -0.0031 &  9.9 & a \\
2001-02-20T09:48:26.003	& -0.6161 &  0.0394 & -0.6294 &  0.0317 &  0.0132 &  0.0077 &  9.9 & a \\
2001-02-23T13:40:11.997	& -0.2532 &  0.1957 & -0.2688 &  0.1861 &  0.0156 &  0.0096 & 40.2 & b \\[0.2cm]

2004-07-25T10:22:42.003	& -0.7657 & -0.0070 & -0.7530 & -0.0084 & -0.0128 &  0.0015 & 13.3 & c \\
2004-08-10T07:19:16.000	&  0.3738 &  0.1392 &  0.3877 &  0.1408 & -0.0139 & -0.0016 & 13.3 & c \\
2004-08-29T07:20:19.003	&  0.7728 & -0.1082 &  0.7916 & -0.0946 & -0.0188 & -0.0136 & 13.3 & c \\
2004-08-29T07:28:17.996	&  0.7940 & -0.1065 &  0.7890 & -0.0963 &  0.0050 & -0.0102 & 13.3 & c \\
2004-08-29T08:48:08.003	&  0.7434 & -0.1190 &  0.7584 & -0.1132 & -0.0150 & -0.0058 & 13.3 & c \\
2004-09-01T05:57:00.996	&  0.5454 &  0.1366 &  0.5514 &  0.1363 & -0.0060 &  0.0003 & 13.3 & c \\
2004-09-01T06:05:55.000	&  0.5545 &  0.1330 &  0.5581 &  0.1346 & -0.0036 & -0.0016 & 13.3 & c \\
2004-09-01T08:27:37.995	&  0.6583 &  0.1065 &  0.6552 &  0.1057 &  0.0031 &  0.0007 & 13.3 & c \\
2004-09-03T07:11:05.999	& -0.7498 & -0.0610 & -0.7497 & -0.0599 & -0.0001 & -0.0011 & 13.3 & c \\
2004-09-04T08:43:35.999	& -0.1603 &  0.1950 & -0.1423 &  0.2135 & -0.0180 & -0.0185 & 13.3 & c \\
2004-09-05T04:11:27.997	&  0.7851 &  0.0490 &  0.7862 &  0.0433 & -0.0012 &  0.0057 & 13.3 & c \\
2004-09-05T08:17:20.002	&  0.8362 & -0.0192 &  0.8363 & -0.0191 & -0.0001 & -0.0002 & 13.3 & c \\
2004-09-06T03:52:03.999	&  0.1661 & -0.2058 &  0.1858 & -0.2126 & -0.0197 &  0.0068 & 13.3 & c \\
2004-09-07T02:38:22.004	& -0.8307 &  0.0053 & -0.8216 &  0.0046 & -0.0091 &  0.0007 & 13.3 & c \\
2004-09-07T09:26:55.999	& -0.7914 &  0.1163 & -0.7785 &  0.1060 & -0.0129 &  0.0104 & 13.3 & c \\
2004-09-08T06:56:51.002	&  0.2542 &  0.1941 &  0.2694 &  0.1926 & -0.0152 &  0.0015 & 13.3 & c \\
2004-09-11T04:53:29.996	& -0.6726 &  0.1650 & -0.6542 &  0.1600 & -0.0184 &  0.0050 & 13.3 & c \\
2004-09-13T03:27:24.001	&  0.5571 & -0.1980 &  0.5477 & -0.1852 &  0.0094 & -0.0128 & 13.3 & c \\
2004-09-13T05:32:11.002	&  0.4478 & -0.1852 &  0.4482 & -0.2004 & -0.0004 &  0.0153 & 13.3 & c \\
2004-09-14T03:51:50.996	& -0.7220 & -0.0725 & -0.7136 & -0.0758 & -0.0084 &  0.0033 & 13.3 & c \\
2004-09-14T06:38:29.999	& -0.7710 & -0.0361 & -0.7811 & -0.0330 &  0.0101 & -0.0031 & 13.3 & c \\
2004-09-14T06:46:34.003	& -0.7829 & -0.0314 & -0.7836 & -0.0309 &  0.0007 & -0.0005 & 13.3 & c \\
2004-09-15T05:18:22.003	& -0.2080 &  0.2215 & -0.2048 &  0.2204 & -0.0032 &  0.0011 & 13.3 & c \\
2004-10-19T00:29:36.003	&  0.7423 & -0.0400 &  0.7424 & -0.0392 & -0.0001 & -0.0008 & 13.3 & c \\
2004-10-20T00:10:57.996	& -0.0891 & -0.1748 & -0.0993 & -0.1846 &  0.0101 &  0.0098 & 13.3 & c \\
2004-10-25T06:29:31.001	& -0.1298 &  0.1851 & -0.1333 &  0.1891 &  0.0034 & -0.0040 &  9.9 & c \\
2004-11-02T07:35:11.999	&  0.6486 &  0.0460 &  0.6243 &  0.0427 &  0.0244 &  0.0033 & 22.0 & c \\[0.2cm]

2005-11-01T13:20:57.001	&  0.0015 & -0.2670 & -0.0046 & -0.2581 &  0.0061 & -0.0089 & 22.0 & e \\
2005-11-06T08:20:06.002	&  0.6840 &  0.1901 &  0.6911 &  0.1814 & -0.0071 &  0.0088 & 22.0 & e \\
2005-11-06T08:32:42.996	&  0.6914 &  0.1739 &  0.6862 &  0.1844 &  0.0051 & -0.0105 & 22.0 & e \\
2005-12-20T10:01:09.001	&  0.2384 &  0.2573 &  0.2450 &  0.2594 & -0.0066 & -0.0021 & 22.0 & e \\
2005-12-20T10:10:25.996	&  0.2434 &  0.2434 &  0.2375 &  0.2596 &  0.0059 & -0.0162 & 22.0 & e \\
2005-12-21T08:42:09.002	& -0.7005 &  0.0065 & -0.6844 &  0.0035 & -0.0161 &  0.0030 & 22.0 & e \\
2005-12-21T08:48:50.002	& -0.6704 &  0.0129 & -0.6860 &  0.0014 &  0.0156 &  0.0115 & 22.0 & e \\[0.2cm]

2006-01-01T10:44:37.003	& -0.6687 & -0.0559 & -0.6808 & -0.0677 &  0.0121 &  0.0118 & 22.0 & e \\
2006-01-06T09:14:37.996	&  0.4125 & -0.1554 &  0.4061 & -0.1544 &  0.0063 & -0.0009 & 22.0 & e \\
2006-01-06T09:22:55.997	&  0.4112 & -0.1418 &  0.4113 & -0.1525 & -0.0000 &  0.0106 & 22.0 & e \\
2006-12-12T16:09:34.001	&  0.5659 &  0.1796 &  0.5353 &  0.1994 &  0.0306 & -0.0199 &  9.9 & e \\[0.2cm]

2010-07-27T05:23:00.040	&  0.3900 &  0.2950 &  0.3719 &  0.3024 &  0.0181 & -0.0074 & 12.5 & e \\
2010-08-15T08:37:51.997	&  0.6981 & -0.2980 &  0.6827 & -0.2803 &  0.0155 & -0.0177 & 22.0 & e \\
2010-08-25T08:30:23.002	&  0.4274 &  0.2690 &  0.4061 &  0.2778 &  0.0213 & -0.0089 & 22.0 & e \\
2010-08-28T08:23:46.996	& -0.4167 &  0.3680 & -0.3989 &  0.3656 & -0.0178 &  0.0024 & 22.0 & e \\
2010-08-30T05:57:28.998	&  0.3225 & -0.3714 &  0.3252 & -0.3766 & -0.0027 &  0.0052 & 13.3 & f \\
2010-09-01T08:29:10.003	&  0.0744 &  0.3535 &  0.0584 &  0.3638 &  0.0161 & -0.0102 & 22.0 & e \\
2010-09-02T06:35:15.002	&  0.7748 & -0.1153 &  0.7493 & -0.1171 &  0.0256 &  0.0018 & 22.0 & e \\[0.2cm]

2011-10-07T02:37:58.002	& -0.6980 & -0.0720 & -0.7047 & -0.0615 &  0.0067 & -0.0105 & 13.3 & d \\
2011-11-06T02:03:15.001	&  0.3680 & -0.0190 &  0.3771 & -0.0250 & -0.0091 &  0.0060 & 13.3 & d \\
2011-11-08T03:05:36.000	& -0.6060 &  0.0150 & -0.6088 &  0.0054 &  0.0028 &  0.0096 & 13.3 & d \\
2011-11-10T01:04:56.000	&  0.6430 &  0.0120 &  0.6741 &  0.0015 & -0.0311 &  0.0105 & 13.3 & d \\
2011-11-15T00:36:51.001	& -0.1770 &  0.0460 & -0.1824 &  0.0373 &  0.0054 &  0.0087 & 13.3 & d \\
2011-11-16T01:07:05.998	& -0.6700 & -0.0450 & -0.6832 & -0.0423 &  0.0132 & -0.0027 & 13.3 & d \\
2011-11-20T01:06:14.002	& -0.3550 & -0.0350 & -0.3609 & -0.0527 &  0.0059 &  0.0177 & 13.3 & d \\
2011-12-15T05:06:33.998	& -0.5680 & -0.0570 & -0.5736 & -0.0483 &  0.0056 & -0.0087 &  9.9 & d \\
2011-12-15T05:29:28.000	& -0.5680 & -0.0630 & -0.5644 & -0.0495 & -0.0036 & -0.0135 &  9.9 & d \\
2011-12-15T06:28:30.002	& -0.5390 & -0.0460 & -0.5387 & -0.0523 & -0.0003 &  0.0063 &  9.9 & d \\
2011-12-16T04:30:17.003	&  0.3910 & -0.0370 &  0.3829 & -0.0392 &  0.0081 &  0.0022 &  9.9 & d \\
2011-12-16T06:01:26.002	&  0.4460 & -0.0160 &  0.4385 & -0.0334 &  0.0075 &  0.0174 &  9.9 & d \\
2011-12-17T06:16:33.003	&  0.4150 &  0.0820 &  0.4080 &  0.0612 &  0.0070 &  0.0208 &  9.9 & d \\
2011-12-18T04:43:58.002	& -0.5280 &  0.0260 & -0.5306 &  0.0200 &  0.0026 &  0.0060 &  9.9 & d \\
2011-12-18T04:53:02.002	& -0.5210 &  0.0190 & -0.5346 &  0.0193 &  0.0136 & -0.0003 &  9.9 & d \\[0.2cm]

2013-01-06T01:38:14.000	&  0.5930 & -0.1130 &  0.5898 & -0.1041 &  0.0032 & -0.0089 &  4.5 & g \\

\end{longtable}
\end{small}

\clearpage

\begin{small}
\begin{longtable}{clrllc}
 \caption{Timings and uncertainties of the occultation of TYC2 1856-00745-1 by (87) Sylvia on January 6, 2013 (source: {\em Euraster} Web site). Notes: $^{a}$ Target: S for Sylvia, R for Romulus, $^{b}$ Measurement uncertainty not provided by the observer, $^{c}$ No absolute timing due to technical problem; the only reliable data is the duration of the event, and the delay between the two disappearance times, which is 97.10 \rev{$\pm 0.03$} sec.} \label{tbl-timings} \\

\hline \noalign{\smallskip}
 T $^{a}$ & Name & Duration & Disp. Time & Reap. Time & Offset \\
  &  & (s) & (h:m:s $\pm$s) & (h:m:s $\pm$s) & (s) \\
\hline \noalign{\smallskip}
\endfirsthead

\multicolumn{6}{c}%
{{\bfseries \tablename\ \thetable{} -- continued}} \\

\hline \noalign{\smallskip}
 T $^{a}$ & Name & Duration & Disp. Time & Reap. Time & Offset \\
  &  & (s) & (h:m:s $\pm$s) & (h:m:s $\pm$s) & (s) \\
\hline \noalign{\smallskip}
\endhead

\hline \noalign{\smallskip}
\multicolumn{6}{r}{{Continued on next page}} \\
\endfoot

\noalign{\smallskip}\hline
\endlastfoot

S & A. Carbognani           & 17.70 & 01:37:33.30 $\pm$0.28   & 01:37:51.00 $\pm$0.28 & \\
S & S. Bolzoni              & 18.20 & 01:37:29.15 $\pm$0.30   & 01:37:47.35 $\pm$0.20 & \\
S & S. Sposetti, A. Manna   & 25.34 & 01:37:26.21 $\pm$0.03   & 01:37:51.55 $\pm$0.04 & \\
S & V. Fristot              & 21.50 & 01:37:30.70 $\pm$0.20   & 01:37:52.20 $\pm$0.20 & \\
S & A. Figer                & 26.20 & 01:37:22.80 $\pm$0.60   & 01:37:49.00 $\pm$0.80 & \\
S & M. Bretton              & 18.10 & 01:37:34.50 $\pm$0.0 \textsuperscript{b} & 01:37:52.60 $\pm$0.0 \textsuperscript{b} & -3.8 \\
S & J. Lecacheux            &  4.06 & 01:37:37.16 $\pm$0.08   & 01:37:41.22 $\pm$0.10 & \\
S & P. Tanga                & 19.19 & \rev{00:00:00.00 $\pm$0.02} \textsuperscript{c} & \rev{00:00:19.19 $\pm$0.02} \textsuperscript{c} & \\
S & M. Devogele et al.      & 12.77 & 01:37:26.73 $\pm$0.08   & 01:37:39.50 $\pm$0.08 & \\
S & E. Frappa, A. Klotz     & 12.90 & 01:37:26.60 $\pm$0.10   & 01:37:39.50 $\pm$0.10 & \\
S & L. Brunetto et al.      &  7.12 & 01:37:29.53 $\pm$0.02   & 01:37:36.65 $\pm$0.02 & \\
S & V. Metallinos           & 24.60 & 01:36:20.73 $\pm$0.05   & 01:36:45.33 $\pm$0.05 & \\
& & & & & \\
R & P. Tanga                &  1.96 & \rev{00:00:97.10 $\pm$0.03} \textsuperscript{c}& \rev{00:00:99.06 $\pm$0.03} \textsuperscript{c} & \\
R & P. Dubreuil             &  1.70 & 01:38:57.60 $\pm$2.00   & 01:38:59.30 $\pm$2.00 & 0.5 \\
R & M. Devogele et al.      &  0.56 & 01:39:00.85 $\pm$0.08   & 01:39:01.41 $\pm$0.08 & \\
R & E. Frappa, A. Klotz     &  0.48 & 01:39:00.73 $\pm$0.08   & 01:39:01.21 $\pm$0.08 & \\

\end{longtable}
\end{small}

\clearpage

\begin{small}
\begin{longtable}{llccrc}
 \caption{Observers of the occultation of TYC2 1856-00745-1 by (87) Sylvia on January 6, 2013 (source: {\em Euraster} Web site). Method is: $<$optic$>$, $<$acquisition$>$, $<$time source$>$, with abbreviations: L: Refractor, M: Reflector, CCD: CCD or CMOS imaging, VID: video recording, VIS: visual, GPS++: GPS one Pulse Per Second, PHONE: Phone time signal (wired phone), RAD: Radio time signal, RAD+: Intermittent radio controlled clock updated just before event, NTP: Network Time Protocol. Coordinates datum: W = WGS84; Altitude datum: W = WGS84, S = Mean Sea Level}  \label{tbl-observers} \\

\hline \noalign{\smallskip}
Name & Method & Longitude & Latitude & Alt. & Datum \\
     &        & (\arcdeg$\,$\arcmin$\,$\arcsec) & (\arcdeg$\,$\arcmin$\,$\arcsec) & (m)  &       \\
\hline \noalign{\smallskip}
\endfirsthead

\multicolumn{6}{c}%
{{\bfseries \tablename\ \thetable{} -- continued}} \\

\hline \noalign{\smallskip}
Name & Method & Longitude & Latitude & Alt. & Datum \\
     &        & (\arcdeg$\,$\arcmin$\,$\arcsec) & (\arcdeg$\,$\arcmin$\,$\arcsec) & (m)  &       \\
\hline \noalign{\smallskip}
\endhead

\hline \noalign{\smallskip}
\multicolumn{6}{r}{{Continued on next page}} \\
\endfoot

\noalign{\smallskip}\hline
\endlastfoot

S. Sposetti           (CH) & M400, VID        & E  09 01 26.5 & N 46 13 53.2 &  260 & WS \\
C. Gualdoni           (IT) & M250, VID        & E  09 06 01.0 & N 45 48 18.0 &  255 & WS \\
A. Carbognani         (IT) & M810, CCD, NTP   & E  07 28 42.0 & N 45 47 23.5 & 1678 & WS \\
S. Bolzoni            (IT) & M300, VIS, RAD+  & E  08 09 59.8 & N 45 21 57.0 &  195 & WS \\
U. Quadri et al.      (IT) & M250, CCD        & E  10 07 49.5 & N 45 19 32.4 &   63 & WS \\
S. Sposetti, A. Manna (IT) & M200, VID, GPS++ & E  07 18 01.3 & N 45 07 34.1 &  585 & WS \\
V. Fristot            (FR) & L102, CCD, NTP   & E  05 16 35.4 & N 44 41 35.5 &  390 & WS \\
A. Figer              (FR) & L68 , CCD, PHONE & E  06 40 54.0 & N 44 34 24.0 & 1845 & WS \\
M. Bretton            (FR) & M820, VIS, RAD   & E  05 30 54.4 & N 44 24 29.9 &  810 & WS \\
J.-L. Penninckx       (FR) & M400, VIS        & E  04 22 45.8 & N 44 07 51.5 &  245 &    \\
C. Peguet             (FR) & M350, VIS        & E  05 00 09.0 & N 44 07 31.8 &   94 & WS \\
L. Bernasconi         (FR) & M500, CCD        & E  05 11 11.2 & N 44 01 17.5 &  330 & WS \\
J. Lecacheux          (FR) & L130, VID, GPS++ & E  06 01 15.3 & N 43 58 48.6 &  720 & WS \\
E. Frappa et al.      (FR) & L61,  VID        & E  05 58 33.4 & N 43 56 07.9 &  460 & WS \\
L. Arnold             (FR) & M205, CCD        & E  05 42 48.4 & N 43 55 00.7 &  551 & WS \\
E. Frappa             (FR) & M203, VID        & E  06 00 23.0 & N 43 51 51.9 &  628 & WS \\
G. Brabant            (FR) & M200, VID        & E  04 54 47.5 & N 43 47 48.8 &   53 & WS \\
P. Tanga              (FR) & M356, CCD        & E  07 15 47.2 & N 43 47 22.2 &  385 & WS \\
P. Dubreuil           (FR) & M203, CCD        & E  07 14 30.6 & N 43 46 58.5 &  480 & WS \\
D. Verilhac           (FR) & M210, VID        & E  04 53 35.1 & N 43 46 01.0 &  115 & WS \\
M. Devogele et al.    (FR) & M1000 ,VID, GPS++& E  06 55 21.8 & N 43 45 13.5 & 1280 & WS \\
E. Frappa, A. Klotz   (FR) & M250, CCD, GPS++ & E  06 55 25.1 & N 43 45 07.3 & 1270 & WS \\
O. Lecacheux          (FR) & L61, VID, GPS++  & E  05 30 00.0 & N 43 41 31.6 &  214 & WS \\
R. Poncy              (FR) & M400, CCD        & E  03 56 24.0 & N 43 38 50.0 &   54 & WS \\
L. Brunetto et al.    (FR) & M406, VID, GPS++ & E  07 04 18.4 & N 43 36 15.7 &  130 & WS \\
D. Albanese           (FR) & M280, CCD        & E  06 39 04.9 & N 43 29 45.9 &   43 & WS \\
F. Colas              (FR) & M1050, CCD       & E  00 08 32.5 & N 42 56 10.9 & 2871 & WS \\
J. Lopez              (ES) & M200, VID        & E  02 59 50.5 & N 42 15 00.2 &   19 & W  \\
C. Perello, A. Selva  (ES) & M500, VID        & E  02 05 24.6 & N 41 33 00.2 &  224 & WS \\
R. Casas              (ES) & M200, VID        & E  02 07 14.3 & N 41 32 22.1 &  165 & WS \\
H. Pallares           (ES) & M280, VIS        & E  04 14 28.1 & N 39 49 12.7 &   19 & W  \\
V. Metallinos         (GR) & L130, VID, GPS++ & E  19 52 23.3 & N 39 38 07.4 &    2 & WS \\

\end{longtable}
\end{small}

\clearpage

\begin{longtable}{lcc}
 \caption{Physical parameters of Sylvia's primary derived using KOALA algorithm.} \label{tbl-physParams} \\

\hline \noalign{\smallskip}
Parameter & Value & Uncertainty (3-$\sigma$) \\
\hline \noalign{\smallskip}
\endfirsthead

\multicolumn{3}{r}%
{{\bfseries \tablename\ \thetable{} -- continued}} \\

\hline \noalign{\smallskip}
Parameter & Value & Uncertainty (3-$\sigma$) \\
\hline \noalign{\smallskip}
\endhead

\hline \noalign{\smallskip}
\multicolumn{3}{r}{{Continued on next page}} \\
\endfoot

\noalign{\smallskip}\hline
\endlastfoot

Spin axis ecliptic J2000 longitude	& 70\arcdeg                 & $\pm 9$\arcdeg \\
Spin axis ecliptic J2000 latitude   & 69\arcdeg                 & $\pm 3$\arcdeg \\
Sidereal rotation period            & 5.183640 h                & $\pm 3 \times 10^{-6}$ h \\
Volume-equivalent diameter \deqV    & 273 km                    & $\pm 10$ km \\
Quadrupole term $J_2$               & 0.024                     & 0.015 -- 0.040 \\
Volume                              & $1.07 \times 10^7$ km$^3$ & $\pm 0.09 \times 10^7$ km$^3$ \\
Surface area                        & $2.47 \times 10^5$ km$^2$ & $\pm 0.15 \times 10^5$ km$^2$ \\
\hline
\multicolumn{3}{c}{Maximum dimensions along the principal axes} \\
$a$                                 & 193 km                    & 180 -- 210 km \\
$b$                                 & 129 km                    & 127 -- 160 km \\
$c$                                 & 124 km                    & 115 -- 150 km \\
\hline
\multicolumn{3}{c}{Dynamically equivalent ellipsoid} \\
$a/c$                               & 1.46                      & 1.30 -- 1.75 \\
$b/c$                               & 1.07                      & 1.02 -- 1.20 \\

\end{longtable}

\clearpage

\begin{longtable}{lrlrl}
 \caption{Remus and Romulus best-fit orbital elements (EQJ2000) obtained using our \emph{Genoid-Kepler} algorithm (orbital period $P$, semi-major axis $a$, eccentricity $e$, inclination $i$, longitude of the ascending node $\Omega$, argument of the pericenter $\varpi$, time of pericenter $t_p$), and derived parameters (primary mass $M$, primary density $\rho$, ecliptic J2000 coordinates of the orbital pole $\lambda_p,\,\beta_p$). Errors are given at 3-$\sigma$.} \label{tbl-dynamicalparam} \\

\hline \noalign{\smallskip}
Element & \multicolumn{2}{c}{Remus} & \multicolumn{2}{c}{Romulus} \\
\hline \noalign{\smallskip}
\endfirsthead

\multicolumn{5}{c}%
{{\bfseries \tablename\ \thetable{} -- continued}} \\

\hline \noalign{\smallskip}
Element & \multicolumn{2}{c}{Remus} & \multicolumn{2}{c}{Romulus} \\
\hline \noalign{\smallskip}
\endhead

\hline \noalign{\smallskip}
\multicolumn{5}{r}{{Continued on next page}} \\
\endfoot

\noalign{\smallskip}\hline
\endlastfoot

  $P$ (day)                   &       1.356654  & $\pm$ $5.9\times10^{-5}$ &        3.641191 & $\pm$ $1.98\times10^{-4}$ \\
  $a$ (km)                    &     684.4       & $\pm$ 109.7              &     1351.7      & $\pm$ 151.0               \\
  $e$                         &       0.0       & $\pm$ 0.05               &        0.007    & $\pm$ 0.047               \\
  $i$ (deg)                   &       8.6       & $\pm$ 13.5               &        8.3      & $\pm$  7.7                \\
  $\Omega$ (deg)              &      93.0       & $\pm$ 48.0               &       93.6      & $\pm$ 20.1                \\
  $\varpi$ (deg)              &     187.9       & $\pm$ 52.1               &      109.1      & $\pm$ 23.6                \\
  $t_{p}$ (JD)                & 2455594.58824   & $\pm$ 0.158              &  2455596.41837  & $\pm$ 0.202               \\
  \hline
  \multicolumn{5}{c}{Derived parameters} \\
  $M$ ($10^{19}$ kg)          &       1.380     & $\pm$ 0.669              &        1.476    & $\pm$ 0.497               \\
  $\rho$ (g.cm$^{-3}$)        &       1.29      & $\pm$ 0.59               &        1.38     & $\pm$ 0.44                \\
  $\lambda_p,\,\beta_p$ (deg) & 70, +65         & $\pm$ 30, 9              & 70, +65         & $\pm$ 17, 4               \\
\end{longtable}

\clearpage

\begin{small}
\begin{longtable}{ccrcrl}
 \caption{Selection of stellar occultations by (87) Sylvia scheduled for the next 10 years. Tycho-2 stars; Mean epoch: approximated time of event; m$_{*}$: magnitude of the target star; $\Delta m$: magnitude drop, $\Delta t$: estimated maximum duration of the event, Location: main area of visibility.} \label{tbl-futurpred} \\

\hline \noalign{\smallskip}
Mean epoch & Star & \multicolumn{1}{c}{m$_{*}$} & \multicolumn{1}{c}{$\Delta m$} & \multicolumn{1}{c}{$\Delta t$} & \multicolumn{1}{c}{Location} \\
(UTC) & TYC2 & mag & mag & (s) &  \\
\hline \noalign{\smallskip}
\endfirsthead

\multicolumn{6}{c}%
{{\bfseries \tablename\ \thetable{} -- continued}} \\

\hline \noalign{\smallskip}
Mean epoch & Star & \multicolumn{1}{c}{m$_{*}$} & \multicolumn{1}{c}{$\Delta m$} & \multicolumn{1}{c}{$\Delta t$} & \multicolumn{1}{c}{Location}  \\
(UTC) & TYC2 & mag & mag & (s) &  \\
\hline \noalign{\smallskip}
\endhead

\hline \noalign{\smallskip}
\multicolumn{5}{r}{{Continued on next page}} \\
\endfoot

\noalign{\smallskip}\hline
\endlastfoot

2016-01-30 07:22 & 6226 01275 & 12.1 & 1.6 &  8.0 & Brazil                                         \\
2016-06-22 23:54 & 6815 03609 & 11.6 & 0.6 & 20.5 & Chile, Argentina, South Africa, Madagascar     \\
2016-08-16 23:36 & 6817 01360 & 12.6 & 1.1 & 44.6 & Chile, Paraguay, Brazil (South)                \\
2016-11-03 23:17 & 6854 00301 & 10.6 & 1.4 &  7.2 & Chile, Argentina, Brazil (South)               \\
2019-10-15 08:11 & 1931 01512 &  9.9 & 2.0 & 14.1 & Chile, Argentina, Brazil (South), South Africa \\
2019-10-20 06:08 & 1932 00479 & 12.0 & 1.9 & 15.8 & Spain, France (South), Italy, Greece           \\
2019-10-29 23:45 & 1932 00469 & 10.0 & 1.9 & 21.0 & Spain, France, Italy, Germany, Poland          \\
2023-07-06 08:01 & 7442 01392 & 11.9 & 1.0 & 22.2 & Mexico, USA (East), Canada (East)              \\

\end{longtable}
\end{small}

\clearpage

\clearpage

\clearpage

\clearpage

\clearpage

\clearpage

\clearpage

\clearpage

\clearpage

\end{document}